\documentclass{article}

\usepackage{arxiv}

\usepackage[utf8]{inputenc}
\usepackage{textcomp}

\usepackage{graphicx}
\usepackage{amsmath}
\usepackage[version=4]{mhchem}
\usepackage{siunitx}
\usepackage{longtable,tabularx}
\usepackage{caption}
\usepackage{subcaption}
\usepackage[export]{adjustbox}
\usepackage{xspace}
\usepackage{hyperref}
\usepackage[numbers]{natbib}

\hypersetup{
pdftitle={Numerical and Experimental Investigation of a NACA 64A-110 Airfoil in Transonic Flow Regime},
pdfauthor={Marcel Blind; Christopher Schauerte; Anne-Marie Schreyer; Andrea Beck},
}

\newcommand{\fref}[1]{Fig.~\ref{#1}}
\newcommand{\tref}[1]{Tab.~\ref{#1}}
\newcommand{\sref}[1]{Sec.~\ref{#1}}
\newcommand{\eref}[1]{Eqn.~(\ref{#1})}

\newcommand*{\cf}{cf.\@\xspace}

\title{Numerical and Experimental Investigation of a NACA 64A-110 Airfoil in Transonic Flow Regime}

\author{Marcel Blind\\Institute of Aerodynamics and Gas Dynamics\\ University of Stuttgart\\ Pfaffenwaldring 21\\ 70569 Stuttgart\\ Germany\\ \texttt{blind@iag.uni-stuttgart.de}\And
Christopher Schauerte\\ Institute of Aerodynamics and Chair of Fluid Mechanics\\ RWTH Aachen University\\ Wüllnerstr. 5a\\ 52062 Aachen\\ Germany \And
Anne-Marie Schreyer\\ Institute of Aerodynamics and Chair of Fluid Mechanics\\ RWTH Aachen University\\ Wüllnerstr. 5a\\ 52062 Aachen\\ Germany \And
Andrea Beck\\Institute of Aerodynamics and Gas Dynamics\\ University of Stuttgart\\ Pfaffenwaldring 21\\ 70569 Stuttgart\\ Germany}
\begin{document}

\maketitle

\begin{abstract}
In this paper we present experimental and numerical reference data for a NACA 64A-110 airfoil at two angles of attack for $Ma=\num{0.72}$ and a Reynolds number of $Re_c=\num{930000}$ with respect to the chord length. The test cases are designed to provide data for an uninclined airfoil at $\SI{0}{\degree}$ and for the case of a stable, steady shock at $\SI{3}{\degree}$. For both cases we conduct experiments in the trisonic wind tunnel at the RWTH Aachen including PIV data as well as numerical wall-resolved large eddy simulations. The results show matching mean velocities and Reynolds stresses in the wake and boundary layer region. The experiment shows a stronger shock including a lambda structure, but the shock location and the overall flow physics are in good agreement for the tested angles of attack. The generated data can be used for code validation, feature development and turbulence modeling.
\end{abstract}

\section*{Nomenclature}

{\renewcommand\arraystretch{1.0}
\noindent\begin{longtable*}{@{}l @{\quad=\quad} l@{}}
$\alpha$       & angle of attack \\
$c$            & chord length \\
$c_a$          & lift coefficient \\
$c_f$          & skin friction coefficient \\
$\Delta_z$     & spanwise extension of the mesh \\
$\vec{f}$      & physical fluxes \\
$Ma$           & Mach number \\
$N$            & Polynomial degree of the numerical scheme \\
$N_\text{geo}$ & geometric order of the mesh \\
$\nu$          & kinematic viscosity \\
$\Omega$       & Reference element \\
$Re_c$         & Reynolds number with respect to the chord length \\
$u$            & solution \\
$u_\infty$     & free stream velocity \\
$u_\tau$       & friction velocity \\
$\overline{u}$, $\overline{v}$ & mean velocities \\
$\overline{u'u'}$, $\overline{v'v'}$, $\overline{u'v'}$ & mean velocity fluctuations \\
$x$, $y$, $z$       & main coordinates \\
$\Delta x$, $\Delta y$, $\Delta z$       & grid spacings \\
$x^+$, $y^+$, $z^+$ & viscous coordinates \\
$\Delta x^+$, $\Delta y^+$, $\Delta z^+$ & viscous grid spacings \\
\end{longtable*}}

\section{Introduction\label{sec:introduction}}

Investigating transonic flow reliably is a challenging task for numerical as well as experimental methods. The presence of physical phenomena such as shocks and separation increase the difficulty of generating comparable data between experiment and simulation. An additional uncertainty results from the wind tunnel itself. Especially transonic flows at lower Reynolds numbers are sensitive towards the geometry of the airfoil, the wind tunnel walls and the blockage ratio. Highly unsteady phenomena such as shock-buffet can further increase the complexity.  In this work we want to study the comparability of experimental data obtained in the trisonic wind tunnel at the RWTH Aachen and numerical simulations. This work aims to provide two reference test cases for validation to act as building blocks for more complex transonic simulations.

Thus, we want to create a reference test case for validation by providing data for a transonic NACA 64A-110 airfoil at two angles of attack.
We designed a combined numerical/experimental study, in order to provide high-fidelity data sets for two representative configurations:
\begin{enumerate}
\item A low angle of attack as a baseline for attached, smooth flow at $\alpha=\SI{0}{\degree}$ and a
\item higher angle of attack scenario at $\alpha=\SI{3}{\degree}$, including a shock.
\end{enumerate}
These cases allow to carefully validate flow solvers, modeling techniques and robustness of the numerical schemes on realistic flow cases. To do so, we conduct experiments in the trisonic wind tunnel at the RWTH Aachen University and use these results to validate wall-resolved large eddy simulation (WLRES) ran with the high-order accurate discontinuous Galerkin spectral element method (DGSEM) framework FLEXI, developed at the University of Stuttgart. The framework FLEXI has been applied to many LES applications in recent years including aero-acoustic simulations and has recently been further developed to account for efficient large eddy simulation \cite{Blind:2022,Kempf:2022}.

The fundamental flow physical parameters of the present test cases include a Reynolds number of $Re_c=\num{930000}$ and a Mach number of $Ma=\num{0.72}$. Especially the $\alpha=\SI{3}{\degree}$ test case is challenging since numerically we have to utilize shock-capturing methods in order to stabilize the numerical scheme, as well as experimentally by suitably adapting the wind tunnel wall contours to enforce the correct Mach number at all given times during the experiment.

The generated particle image velocimetry (PIV) data are then compared with the numerical results of the WRLES including mean velocities and Reynolds stresses. Thus, we can efficiently compare the dimensions of the wake, the location of the shock as well as the velocities in the low pressure area on the suction side of the airfoil as well as parts of the boundary layer profiles on the surface of the airfoil.

In order to generate the reference data, we describe the numerical and experimental parameters used and discuss their suitability. \section{Methodology\label{sec:mehodology}}

For investigation we chose a NACA 6-series airfoil. It was meant to be representative of a typical horizontal tail plane airfoil in terms of thickness distribution and curvature. Additionally, it has the same point of operation. The NACA 64A-110 airfoil thus turned out to be a good choice, delivering all the aspects described above whiles also having a considerable amount of lift ($c_a=0.4-0.5$). The 6A-series cambered airfoils deviate from the typical 6-series NACA airfoil by modifying the chamber-line equation. Thus, the slope of the mean line is held constant from about $x/c=0.85$ to the trailing edge.

In this section we assess the used numerical and experimental methods to generate the data. We start by discussing the numerical scheme and continue by describing the experimental setup.

\subsection{Numerical Method}

To run the wall-resolved simulation we use the open source discontinuous Galerkin spectral element method (DGSEM) framework FLEXI\footnote{\url{https://github.com/flexi-framework/flexi}}. It is developed by the numerics research group at the Institute of Aerodynamics and Gas Dynamics at the University of Stuttgart \cite{Krais:2021}. In this work we numerically solve the compressible Navier--Stokes equation using the dimensionless power law for handling viscosity. The compressible Navier--Stokes equations can, as any conservation equation, be written in their common form
\begin{equation}
u_t+\nabla_x \cdot \vec{f}(u,\nabla_x u) = 0,
\label{eqn:conseqn}
\end{equation}
with $u$ denoting the solution vector and $\vec{f}$ the physical fluxes of the equation system. We define the flux $\vec{f}$ as the difference between convective and viscous fluxes $\vec{f}= \vec{f}^c(u)-\vec{f}^v(u,\nabla_x u)$, where the viscous fluxes also depend on the gradient of the solution vector $u$.

In order to discretize the domain, we divide it into an unstructured hexahedral mesh with non-overlapping elements in their reference space $\Omega=[-1,1]^3$ based on the tensorproduct of three one-dimensional Lagrange polynomials of order $N$. We perform the Galerkin method by multiplying \eqref{eqn:conseqn} using the same three-dimensional tensorproduct polynomial as test function and integrating over the reference element $\Omega$.

After introducing the fundamental numerics, we continue assessing the used numerical parameters in order to run the wall-resolved LES simulation. Thus, we have to introduce the approximation of the gradients, the numerical fluxes, the time discretization and the shock-capturing as well as describe the general numerical discretization method of the spatial operations.

Each element in the simulation consists of a $N=7$ basis containing 8 solution points in each dimension of space. Thus, one element consists of $8\times8\times8=512$ degrees of freedom (DOFs). This specific polynomial degree was chosen in order to utilize the strengths of a high order scheme, while still providing good domain decomposition capabilities by only having $\num{512}$ DOFs per element. FLEXI has shown to work most efficient by using $\num{3600}$ DOF per process and thus approximately 7 elements are contained in one decomposed domain \cite{Blind:2022}. We evaluate the polynomials at the Legendre--Gauss--Lobatto pointset in order to use kinetic energy preserving split form fluxes according to Pirozzoli \cite{Pirozzoli:2010}. The viscous fluxes, which depend on the spatial gradient of the solution are calculated using the first method of Bassi and Rebay (BR1) \cite{Bassi:1997}.  Additionally, we use a fourth order accurate explicit Runge--Kutta method of Niegemann for advancing the resulting ordinary differential equation in time numerically \cite{Niegemann:2012}. High-order schemes are prone to oscillations and even though we use skew-symmetric split forms to stabilize the scheme, capturing shock poses a significant challenge to the simulation of transonic airfoils \cite{Gassner:2018,Flad:2016}. Thus, we use a sub-cell wise blending approach by \cite{Rueda-Ramirez:2022} to stabilize the scheme. This approach blends the DG solution in each element with a finite volume (FV) solution resulting in a stable scheme even for strong shocks. The blending is done by weighting the element-wise solution between DG and FV for troubled cells. Troubled cells are detected using a Persson shock indicator \cite{Persson:2012}. Thus, the scheme can range from $100\%$ DG to $100\%$ FV in each element. The indicator was tuned to only capture the shock and not to flag elements in the boundary layer as troubled. Additionally, a positivity preserving limiter according to Shu is used to further stabilize the simulation \cite{Zhang:2010}. Especially during the transient period of initializing the flow from the initial free stream state this approach helps stabilizing the scheme further. It was ensured that this limiter was only used during the transient phase of the simulation.

\subsection{Experimental Setup}
\label{sec:ExpApp}

All experiments of this study were carried out in the closed-circuit Trisonic Wind Tunnel facility at the Institute of Aerodynamics, RWTH Aachen University.

\subsubsection{Wind-Tunnel Facility}
The Trisonic Wind Tunnel is an intermittently operated vacuum-type indraft facility that can provide Mach numbers between $0.3 < Ma < 4.0$ with interchangeable test sections and adjustable nozzle geometries. Continuously run screw-type compressors evacuate a vacuum chamber with a total volume of $\SI{380}{\metre\cubed}$ downstream of the test section and transport the air through a dryer bed before it enters a $\SI{180}{\metre\cubed}$ air reservoir at ambient conditions. Turnaround times of approximately 7 minutes are achieved. We keep the relative humidity of the air below 4\% to preclude condensation effects and falsification of the measured shock location \cite{Binion1988}. Upon initialization of a measurement cycle, a fast-acting valve is actuated and the measurement medium (predried air) is sucked from the reservoir through the test section into the vacuum tanks. An effective acquisition phase of stable flow conditions of 2 to 3 seconds is obtained in each measurement cycle. The turbulence intensity in the free stream is below 1\%.

\begin{figure}[t]
\centering
\includegraphics[width=.5\textwidth]{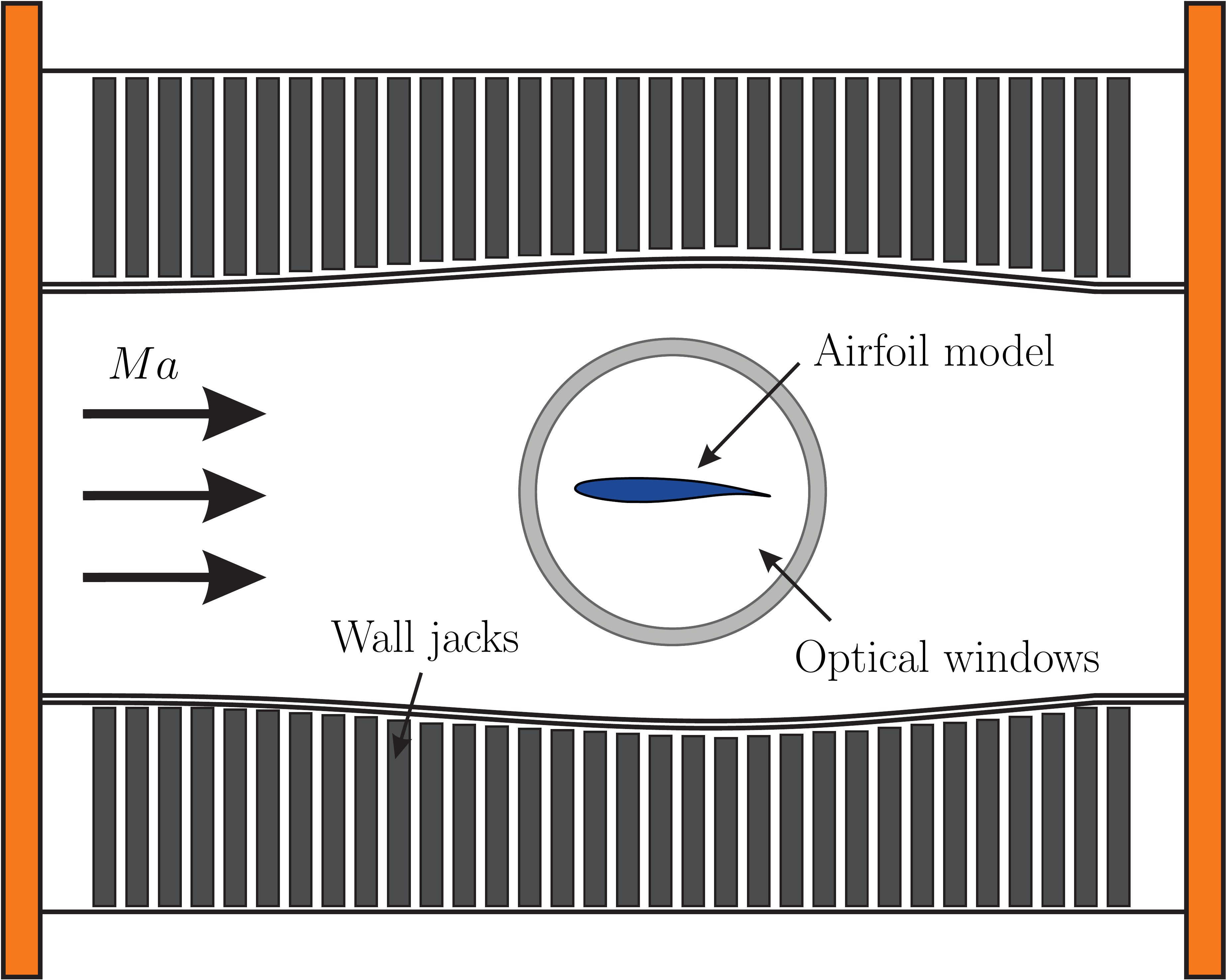}
\caption{Illustration of the transonic test section with adjustable walls and optical access windows.}
\label{fig:wind_tunnel}
\end{figure}

Measurements of the present campaign are carried out in the transonic operation mode. The transonic test section features a rectangular cross section of $\SI{400}{\milli\metre}\times\SI{400}{\milli\metre}$ and a total length of $\SI{1410}{\milli\metre}$, and is equipped with adaptive upper and lower walls to account for influences of solid walls confining the flow and to emulate conditions in terms of streamline contouring observed in the equivalent free, unconfined flow around an airfoil \cite{McDevitt1982}. The adaptation process is based on the transonic small perturbation theory. Wall pressure measurements acquired during each measurement cycle and wall displacement coordinates are used to compute the disturbance velocities by solving the Cauchy’s integral formula \cite{Amecke1985}. Details on the implementation and validation of the method in the trisonic wind tunnel are reported by Romberg \cite{Romberg1990}. Circular windows with a diameter of $\SI{280}{\milli\metre}$ on both sides provide optical access to the test section. A schematic representation of the airfoil installed in the test section is provided in Figure \ref{fig:wind_tunnel}. The obtainable Reynolds number is a function of the selected Mach number and the ambient conditions of the resting, predried air in the reservoir. For the present test campaign, $Ma=0.72$ was chosen, equivalent to a free stream velocity of $u_{\infty}=\SI{235}{\metre\per\second}$. Based on the chord length of the installed NACA 64A-110 airfoil of $c=\SI{75}{\milli\metre}$, a chord-based Reynolds number of $Re_c = \num{930000}$ was obtained.

\subsubsection{Airfoil Model} We examined the transonic flow past the NACA 64A-110 airfoil. The two-dimensional wind tunnel model has a total span of \SI{399}{\milli\metre}, which is equivalent to the width of the test section. With the chord length of $c=\SI{75}{\milli\metre}$, the model aspect ratio is \num{5.32}. To allow for both a high degree of contour accuracy and stiffness, the outer shells were manufactured from metal-particle-filled epoxy resin that was cast around a rigid core consisting of a high-strength stainless steel shaft. The core further allows the rigid installation of the model: the model is installed in the test section via a support through the airfoil core, which is attached to solid steel blocks on either side of the wind tunnel. The support blocks are stiffly connected to the window frames. The whole assembly, including the windows, can be rotated as an entity to adjust the angle of attack ($\alpha$, AoA). The center of rotation for the AoA adjustment is located at about mid-chord of the airfoil.

\subsection{Applied Measurement Techniques}
\label{sec:ExpTechniques}

To characterize the global flow topology, involving the formation of shock waves and the evolution of the airfoil near-wake domain, as well as turbulence, we applied a high-speed focusing schlieren setup and a Particle Image Velocimetry (PIV) arrangement with high spatial resolution.

\subsubsection{High-Speed Focusing Schlieren}
\label{sec:FocSchlieren}

Flow visualizations are obtained with a focusing schlieren setup, where extended grids illuminated by correspondingly large illuminated surfaces replace the point-shaped light sources used in classical schlieren arrangements \cite{Schardin1942,Settles2001}. Schauerte and Schreyer \cite{Schauerte_Schreyer2018} designed a focusing schlieren system inspired by Weinstein's approach \cite{Weinstein1993} to explore flow configurations in the transonic and supersonic Mach number regime. This system overcomes some of the limitations of classical schlieren configurations, in particular the close to infinite depth of focus that results in strong line-of-sight integration. The system allows to focus on narrow slices of the flow, which enables us to analyze three-dimensional flow configurations.

To temporally resolve the shock-induced turbulent flow field, schlieren visualizations were recorded with a Photron SA-5 CMOS high-speed camera at a frame rate of \SI{20}{\kilo\hertz} and with a short exposure of \SI{1.9}{\micro\second}. The flow was illuminated by a $3\times3$ array of \SI{15}{\watt} high-power LEDs together with a constant-current source with an electric output adjustable between \SI{50}{\watt} and \SI{120}{\watt}. The source grid consists of CNC-machined alternating clear and opaque lines, the cutoff grid is an exact photographic negative of the source grid. A large \SI{470}{\milli\metre} Fresnel lens ahead of the source grid captures major fractions of the light intensity, illuminates the source grid evenly by reshaping the light cone, and redirects the light beam through the test section past the model and onto the schlieren lens. The illuminated measurement area around the spanwise center plane of the airfoil has a diameter of \SI{280}{\milli\metre}. However, the outer edges are subject to vignetting due to the light beam passing through two consecutive circular windows. An effective field of view (FOV) without vignetting of $d_{FS} \approx \SI{220}{\milli\metre}$ was achieved, using a magnification $M_{FS}$ of \num{0.2} between the two grids in conjunction with the relay optics and a \SI{35}{\milli\metre} camera lens. For the high-speed recording, a FOV of \SI{195}{\milli\metre} width and  \SI{44}{\milli\metre} height is projected on the camera sensor, yielding an effective resolution of \SI{3.61}{px\per\milli\metre}.

\subsubsection{PIV arrangement}
\begin{figure}[tb!]
\centering
\begin{subfigure}[t]{0.48\textwidth}
\centering
\includegraphics{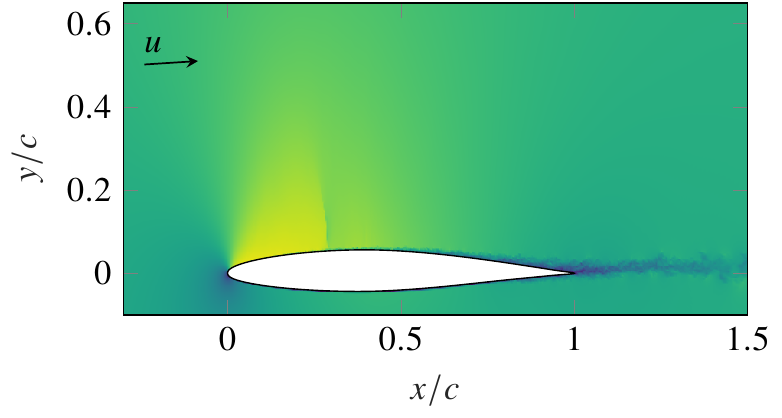}
\caption{Investigated flow field of the WRLES.}
\label{fig:AoA3p0_sim_inst}
\end{subfigure}
\hfill
\begin{subfigure}[t]{0.48\textwidth}
\centering
\includegraphics{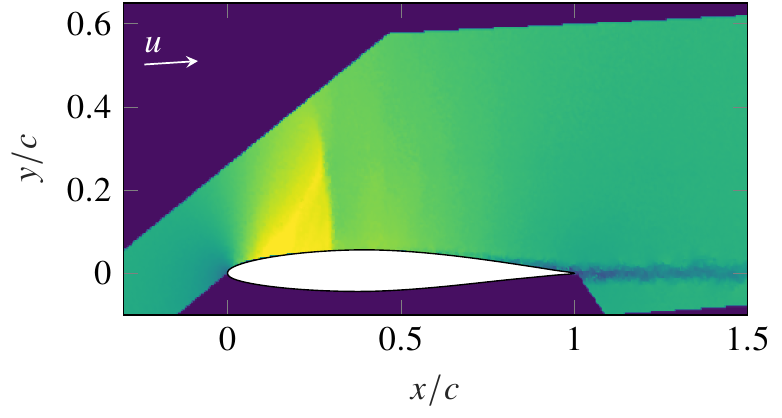}
\caption{Stream-wise velocity contour from PIV. Dark areas indicate the boundaries of the laser light sheet and captured field of view.}
\label{fig:AoA3p0_exp_inst}
\end{subfigure}
\includegraphics{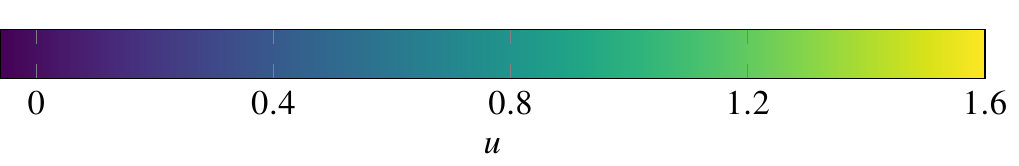}
\caption{Instantaneous snapshot of the u-velocity component showing the $\alpha=\SI{3}{\degree}$ test case.}
\label{fig:AoA3p0_comp_inst}
\end{figure}

To acquire detailed velocity data and allow for quantitative comparisons with the CFD results,
we applied a particle image velocimetry (PIV) system. Planar PIV measurements were carried out in a streamwise/wall-normal plane coinciding with the centerline of the airfoil model. A Litron Nano PIV double pulsed Nd:YLF laser, operating at a wavelength of \SI{527}{\nano\metre}, produces a light sheet of \SI{1.5}{\milli\metre} thickness. The large laser-pulse energy of \SI{200}{\milli\joule\per pulse} with a pulse duration of \SI{4}{\nano\second} ensures a high signal-to-noise ratio. The flow is seeded using di-ethyl-hexyl-sebacate (DEHS) tracer droplets with a mean diameter of \SI{1}{\micro\metre}. Particle images are captured with two identical FlowSense EO 11M CCD cameras with a resolution of \num{11} megapixels, which corresponds to a sensor resolution of $\SI{4008}{pixels} \times \SI{2672}{pixels}$ at a pixel size of \SI{9}{\micro\metre}. The cameras were focused on the same light sheet and field of view (FoV) and triggered suitably to increase the overall acquisition rate to \SI{10}{\hertz}. The FoV is \SI{142}{\milli\metre} $\times$ \SI{58}{\milli\metre}, thus spanning a domain of $-0.2c$ to $1.5c$ in the chordwise direction, which covers the incoming flow, the entire suction side, as well as the near-wake domain. Based on a magnification of \num{4.12}, a spatial resolution of \SI{28.22}{px \per \milli\metre} was obtained. Relevant parameters of the focusing schlieren and PIV tests are summarized in table \ref{tab:ExperimentalParameters}. A comparison between the experimental PIV FoV and the corresponding numerical region for the $\alpha=\SI{3}{\degree}$ test case can be found in \fref{fig:AoA3p0_comp_inst}.

\begin{table}
\caption{Overview of experimental PIV and FS parameters.}
\label{tab:ExperimentalParameters}
\centering
\begin{tabular}{lllllll}
\hline
\multicolumn{7}{l}{Particle image velocimetry setup\hspace{2.1cm} ~ Focusing schlieren setup}\\
\cline{1-3} \cline{4-7}
Sensor resolution & \num{4008} $\times$ \num{2672} & \si{px} $\times$ \si{px} & & Sensor resolution & \num{704} $\times$ \num{520} & \si{px} $\times$ \si{px} \\
FOV width & \num{142} & \si{\milli\metre} & & FOV width & \num{195} & \si{\milli\metre} \\
FOV height & \num{58} & \si{\milli\metre} & & FOV height & \num{144} & \si{\milli\metre} \\
Acquisition rate & \num{10} & \si{\hertz} & & Acquisition rate & \num{20000} & \si{\hertz} \\
Exposure & \num{11} & \si{\micro\second} & & Shutter speed & 1/\num{525000} & \si{\second} \\
Spatial resolution & \num{28.22} & \si{px\per\milli\metre} & & Spatial resolution & \num{3.61} & \si{px\per\milli\metre} \\
\hline
\end{tabular}
\end{table}

 \section{Test Cases\label{sec:testcases}}

In course of this paper, we investigate the flow of the NACA 64A-110 airfoil at two angles of attack. An uninclined $\alpha=\SI{0}{\degree}$ test case as well as a $\alpha=\SI{3}{\degree}$ test case containing a steady shock. Throughout the setup of numerical simulations and the experiments we aim to ensure comparability. Thus, we now discuss how we set up the experiments and the simulations.

\subsection{Wind Tunnel Walls}

The trisonic wind tunnel has adaptive walls (see also \fref{fig:wind_tunnel}) in order to account for influences of solid walls confining the flow around the airfoil. Due to the relatively low Reynolds number and the resulting sensitivity of the flow, we chose to mimic the walls in the numerical simulation.

\begin{figure}[tb!]
\centering
\begin{subfigure}[t]{0.48\textwidth}
\centering
\includegraphics{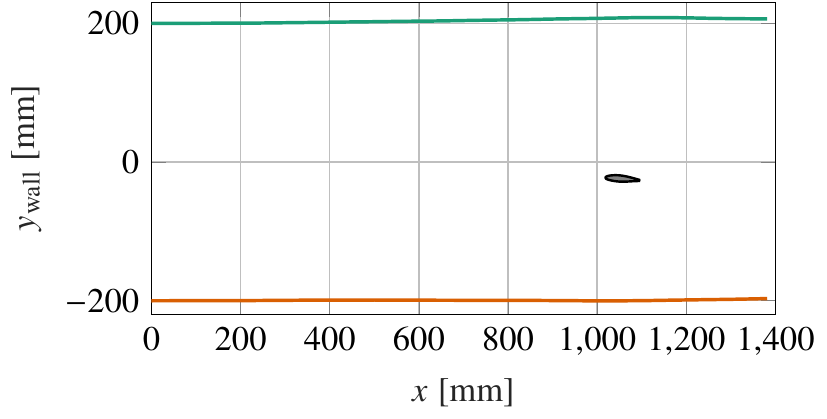}	
\caption{Overview of the NACA airfoil in the wind tunnel with modified walls.}
\label{fig:wall_overview}
\end{subfigure}
\begin{subfigure}[t]{0.48\textwidth}
\centering
\includegraphics{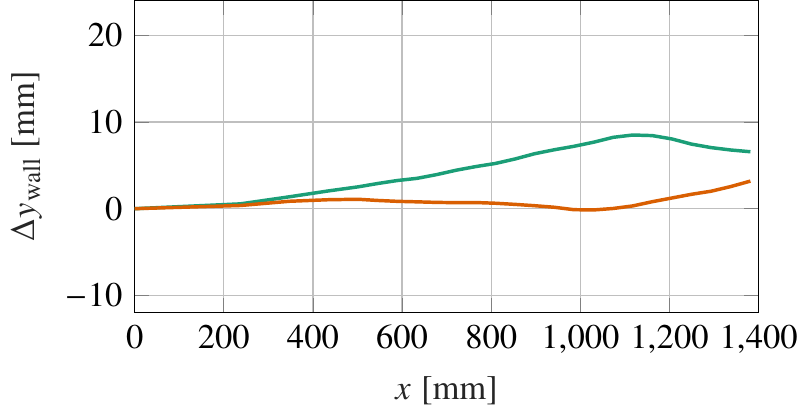}
\caption{Detailed view of change in the wall coordinates.}
\label{fig:wall_delta}
\end{subfigure}
\includegraphics{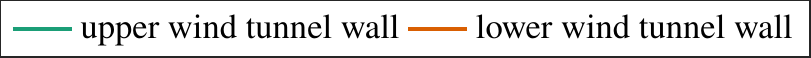}
\caption{Wind tunnel wall contours for $\alpha=\SI{3}{\degree}$ showing the position of the NACA 64A-110 airfoil in the tunnel as well as the detailed change of wind tunnel geometry needed to match the desired flow parameters.}
\label{fig:walls}
\end{figure}

For the $\alpha=\SI{0}{\degree}$ there was barely any change of the wall geometry necessary in the wind tunnel. In simulation we thus modeled the walls straight without any curvature. For $\alpha=\SI{3}{\degree}$ the actual change in shape is visualized in \fref{fig:walls}. We noticed that small changes to the wind tunnel geometry can have significant influence on the position of the occurring shock. Therefore, we applied the exact same wall displacement in \fref{fig:wall_delta} also to the computational grid.

The wind tunnel walls are modeled as Euler walls. This is to avoid the additional computational expense of resolving a boundary layer above and below the airfoil. Additionally, the thickness of the incoming wall boundary layers is unknown and thus cannot be applied via a common turbulent inflow boundary condition.

\subsection{Laminar - Turbulent Transition}

\begin{figure}[tb!]
\centering
\begin{subfigure}[t]{0.475\textwidth}
\centering
\includegraphics[width=0.9\columnwidth]{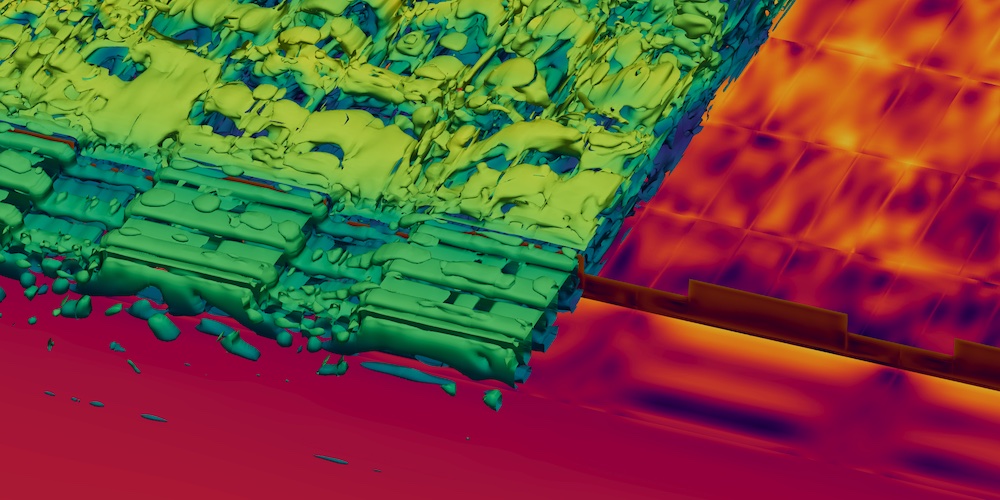}
\caption{Geometric trip forcing transition to turbulence at the $\alpha=\SI{0}{\degree}$ case.}
\label{fig:trip_geo}
\end{subfigure}
\hfill
\begin{subfigure}[t]{0.475\textwidth}
\centering
\includegraphics[width=0.9\columnwidth]{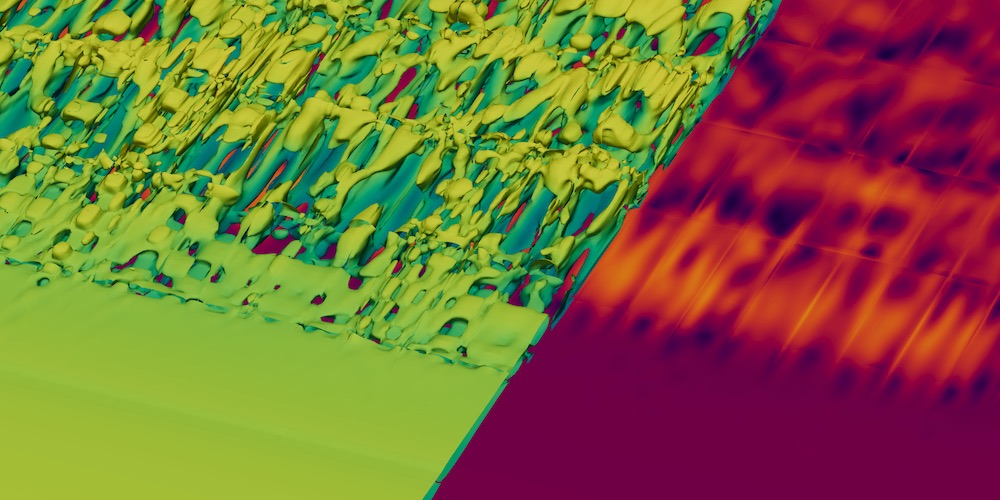}
\caption{Numerical trip forcing transition to turbulence at the $\alpha=\SI{3}{\degree}$ case.}
\label{fig:trip_num}
\end{subfigure}
\caption{Comparison of different trip mechanisms used in this study to force transition to turbulent flow. Surface color shows wall skin friction magnitude. Turbulent structures are visualized using an iso-surface of the Q-Criterion on the left part of each figure. The color of the iso-surfaces depicts the streamwise velocity component.}
\label{fig:trips}
\end{figure}

Another major influence on the flow over the airfoil is the forced transition to turbulent flow. For the WRLES two approaches have been used. In case of the $\alpha=\SI{0}{\degree}$, we utilized a geometrical trip at $x/c=0.05$. The trip has a three-dimensional pattern to improve transition and had an extension up to around $y^+\in[40,50]$ in the boundary layer on the suction and pressure side. The area where the trip is active is visualized in \fref{fig:trip_geo}. Since creating a geometric trip involves remeshing and modifications hence are hard to realize, we chose to also validate a numerical trip proposed by Schlatter and Örlu \cite{Schlatter:2012}. The transition is achieved by applying a volume forcing in the corresponding boundary layer elements as visualized in \fref{fig:trip_num}. Again, the trip is placed at $x/c=0.05$ on the pressure and suction side of the airfoil.

Comparing the simulation tripping methods in \fref{fig:trips} we can see that the numerical trip produces less overall noise and shows fewer artifacts upstream. We note that due to the relatively low Reynolds number, reliable transition to turbulence could not be achieved on the pressure side of the $\alpha=\SI{3}{\degree}$ inclined airfoil. Neither geometric nor numerical trip could achieve this without causing instabilities due to nonphysical oscillations.

In the experiment on the other hand the transition was achieved by a fixed \SI{120}{\micro\metre} zig-zag strips on both sides of the airfoil at a streamwise location of $x/c=0.05$. If one converts the height of the geometric trip in the $\alpha=\SI{0}{\degree}$ test case to match the experimental boundary conditions we obtain a physical height of \SI{95}{\micro\metre} of the geometric trip and a viscous height of the experimental trip of $y^+\approx63$ respectively.

\subsection{Meshes}

\begin{figure}[tb!]
\centering
\begin{subfigure}[t]{0.48\textwidth}
\centering
\includegraphics[width=0.9\columnwidth,frame]{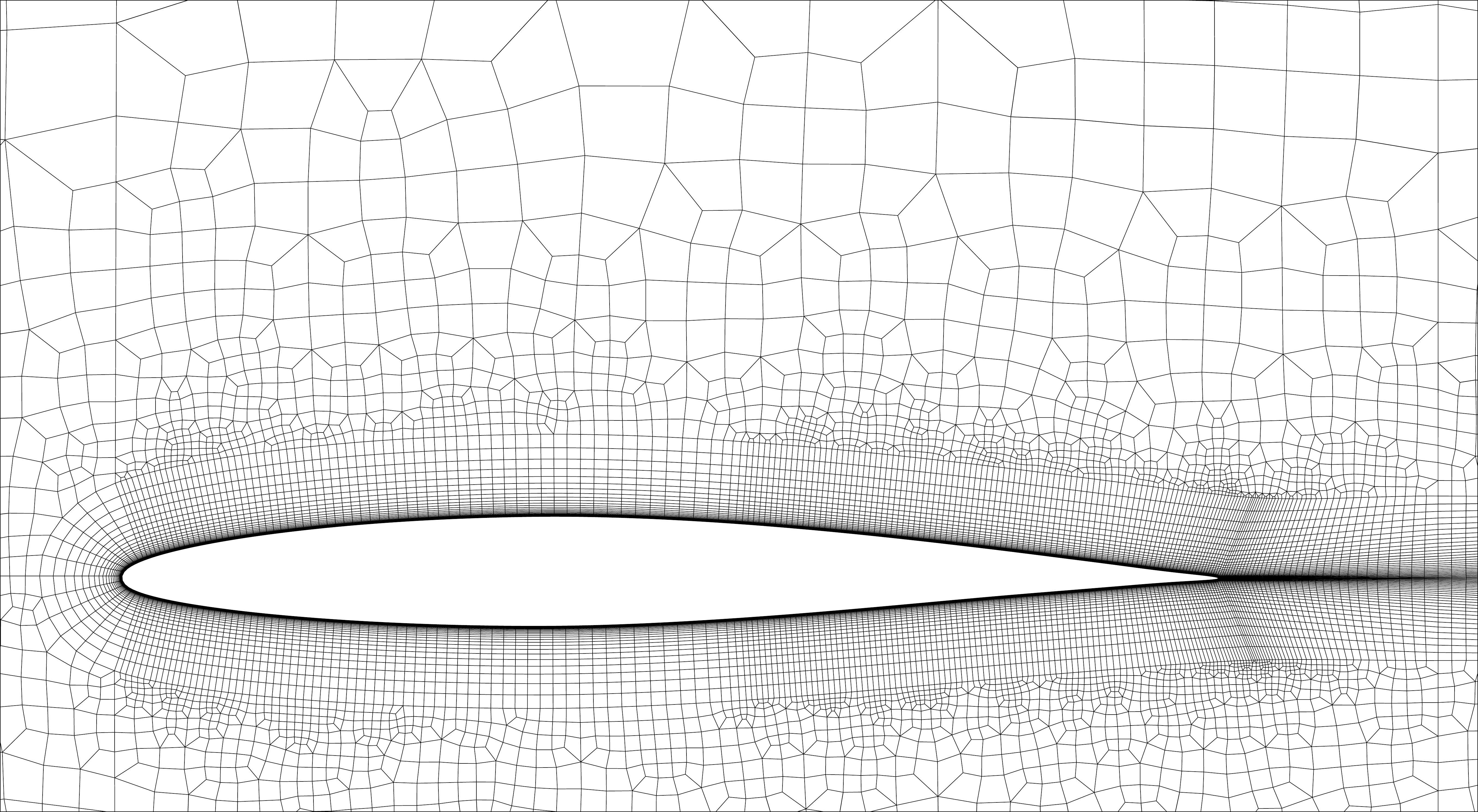}
\caption{Mesh used for the $\alpha=\SI{0}{\degree}$ simulation.}
\label{fig:mesh_AoA0p0}
\end{subfigure}
\hfill
\begin{subfigure}[t]{0.48\textwidth}
\centering
\includegraphics[width=0.9\columnwidth,frame]{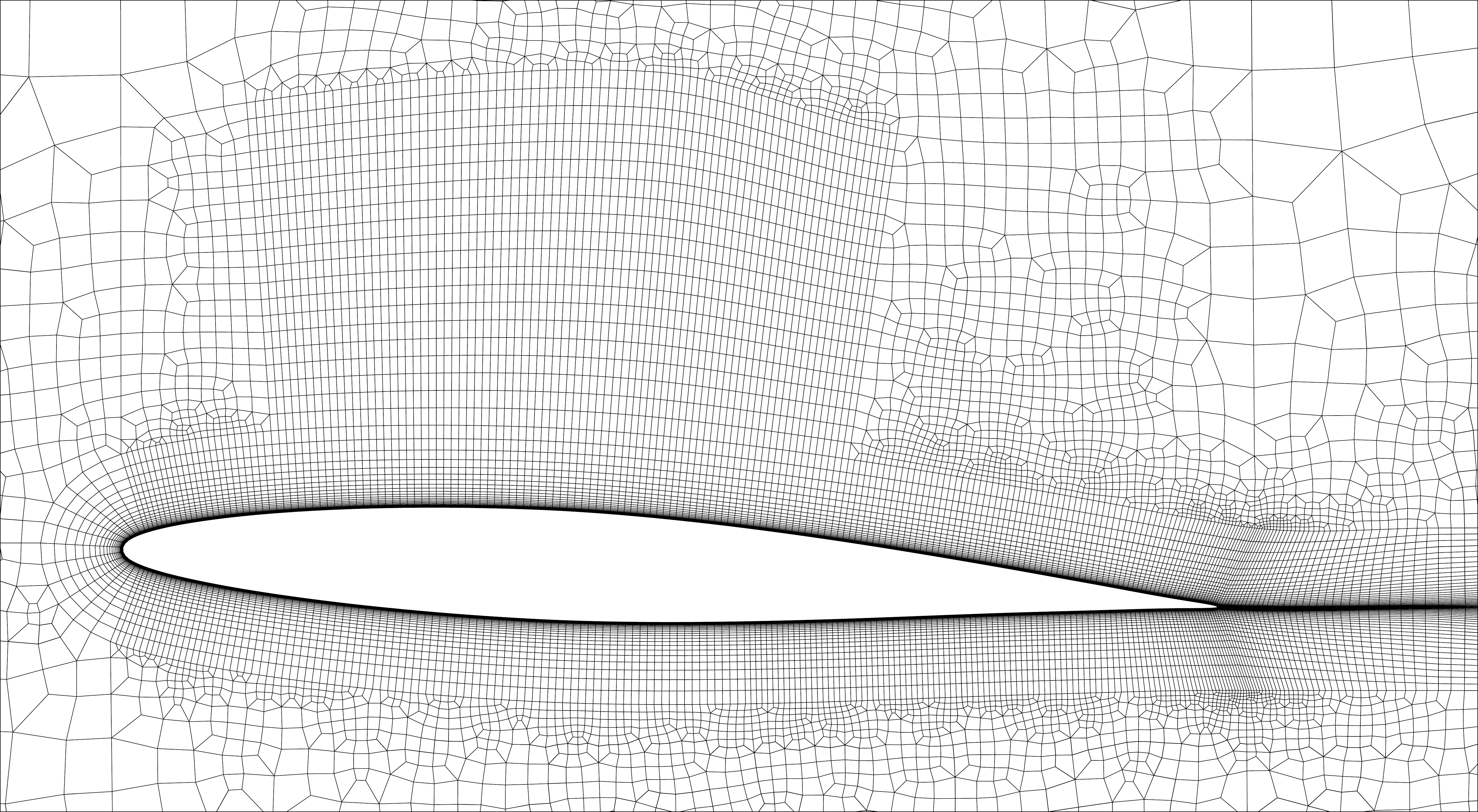}
\caption{Mesh used for the $\alpha=\SI{3}{\degree}$ simulation including refined area on the suction side for the expected shock.}
\label{fig:mes_AoA3p0}
\end{subfigure}
\caption{Comparison of the meshes used to run the wall-resolved large eddy simulations.}
\label{fig:meshes}
\end{figure}

To run the simulations, we created two meshes. This was necessary, since the wind tunnel wall was different for both angle of attacks. Additionally, we had to incline the airfoil $\alpha=\SI{3}{\degree}$ in order to correctly mimic the wind tunnel conditions. The $\alpha=\SI{3}{\degree}$ mesh also has an additional refinement area on the suction side for capturing the shock correctly. Both meshes are visualized in \fref{fig:meshes}.

Both meshes are designed to fulfill the common WRLES criteria as proposed by \cite{Wagner:2009,Georgiadis:2010} and are designed to meet the criteria

\begin{equation}
\Delta x^+<55,\quad \Delta y^+_\text{wall}<1 \quad \text{and} \quad \Delta z^+<10.
\label{eqn:criteria}
\end{equation}

For creating the meshes we used the integral method proposed by Eppler \cite{Eppler:1963,Eppler:1980}. With
\begin{equation}
x^+=\dfrac{u_\tau\Delta x}{\nu} \quad\Longrightarrow\quad \Delta x = \dfrac{x^+(N+1)\nu}{u_\tau}=\dfrac{x^+(N+1)}{Re\sqrt{\frac{c_f}{2}}}
\label{eqn:spacing}
\end{equation}
in case of dimensionless simulation we obtain the necessary grid spacing $\Delta x$ according to the chosen resolution criteria in \eref{eqn:criteria}. The factor $(N+1)$ in \eref{eqn:spacing} accounts for the solution points in each element and $c_f$ the skin friction coefficient as obtained by the integral method. Calculation of $y^+$ and $z^+$ is done equivalently. The spacing in $y$ and $z$ direction is fixed in the C-mesh used for meshing the airfoils. Only the wall-parallel spacing $\Delta x$ is varied as a function of $x/c$ to save elements.

\begin{figure}[tb!]
\centering
\begin{subfigure}[t]{0.48\textwidth}
\centering
\includegraphics{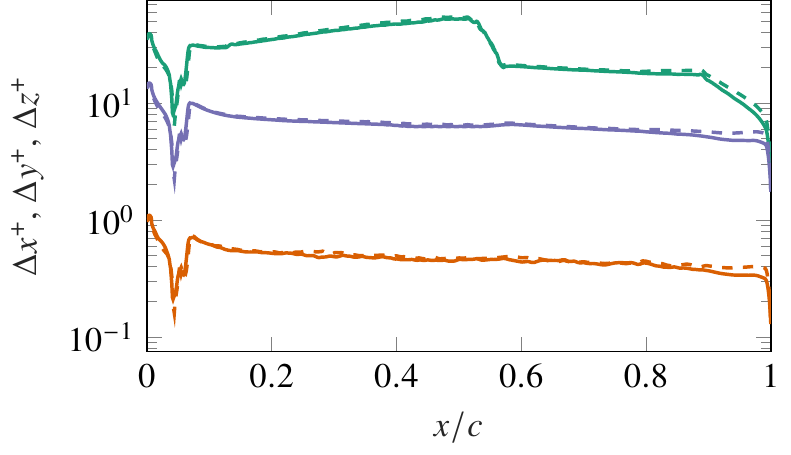}
\caption{Viscous wall spacings for the $\alpha=\SI{0}{\degree}$ simulation.}
\label{fig:res_AoA0p0}
\end{subfigure}
\hfill
\begin{subfigure}[t]{0.48\textwidth}
\centering
\includegraphics{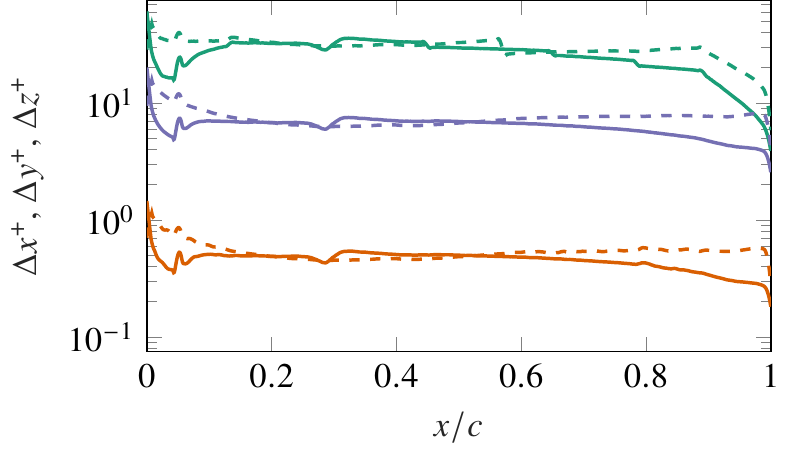}
\caption{Viscous wall spacings for the $\alpha=\SI{3}{\degree}$ simulation.}
\label{fig:res_AoA3p0}
\end{subfigure}
\includegraphics{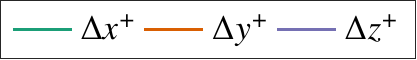}
\hspace{0.05\columnwidth}
\includegraphics{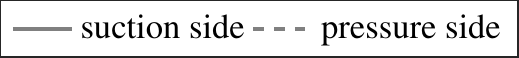}
\caption{Comparison of the viscous wall spacings in the firs off-wall element for the wall-resolved large eddy simulations.}
\label{fig:resolution}
\end{figure}

To check the mesh quality, we analyzed the viscous wall units in the first off-wall element after running the wall-resolved LES. The results are visualized in \fref{fig:resolution}.

\begin{figure}[tb!]
\centering
\begin{subfigure}[t]{0.48\textwidth}
\centering
\includegraphics{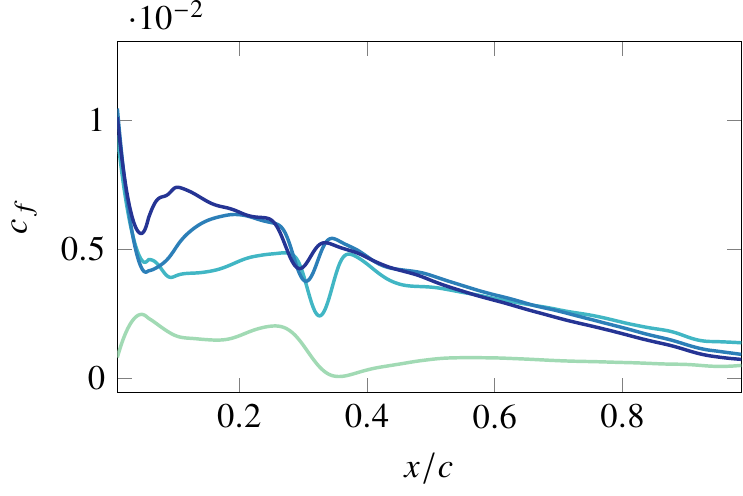}
\caption{Convergence of wall skin friction $c_f$ for the $\alpha=\SI{3}{\degree}$ simulation with different polynomial degree $N$.}
\label{fig:convergence_cf}
\end{subfigure}
\hfill
\begin{subfigure}[t]{0.48\textwidth}
\centering
\includegraphics{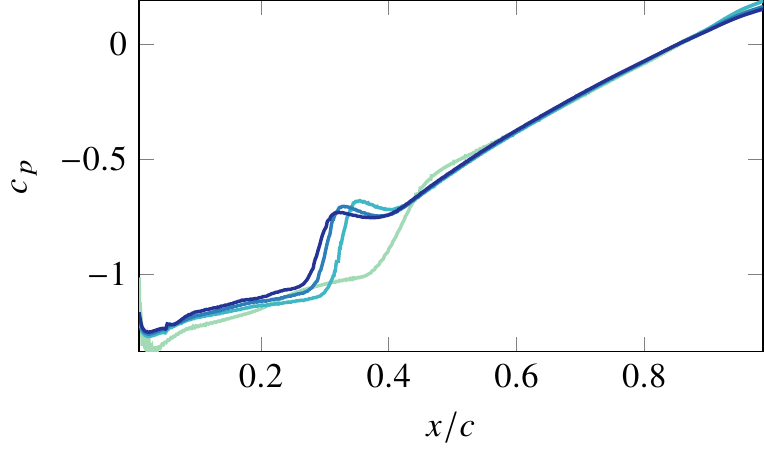}
\caption{Convergence of pressure coefficient $c_p$ for the $\alpha=\SI{3}{\degree}$ simulation with different polynomial degree $N$}
\label{fig:convergence_cp}
\end{subfigure}
\includegraphics{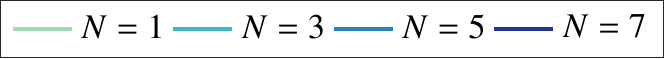}
\caption{Convergence of the integral flow quantities $c_f$ and $c_p$ on the suction side of the airfoil for increasing polynomial degree $N$ and an angle of attack of $\alpha=\SI{3}{\degree}$.}
\label{fig:convergence}
\end{figure}

We can see that the criteria chosen in \eref{eqn:criteria} are all fulfilled by the simulation for $\alpha=\SI{0}{\degree}$ and $\alpha=\SI{3}{\degree}$. For \fref{fig:mesh_AoA0p0} we see an increase of $\Delta x^+$ towards $x/c=0.5$. This was done to save elements in the area with low pressure gradient. The resolution is again refined towards the trailing edge. The overall resolution in $x$ has increased for the $\alpha=\SI{3}{\degree}$ test case in order to also be able to capture the shock better. We show convergence for the integral flow quantities for the $\alpha=\SI{3}{\degree}$ test case in \fref{fig:convergence}. Convergence is reached by increasing the polynomial degree $N$ of the simulation starting at $N=1$.

The pressure coefficient $c_p$ in \fref{fig:convergence_cp} reaches convergence very fast. Only the shock position requires the higher polynomial degree to reach a steady location. In \fref{fig:convergence_cf} the convergence of the skin friction coefficient $c_f$ is shown. This quantity only shows convergence starting $N=3$ due to the dependence on the wall-normal gradient. Convergence for the $\alpha=\SI{0}{\degree}$ test case shows equivalent behavior for these polynomial degrees.

The meshes use high-order elements with a geometric order of $N_\text{geo} = 4$. The resulting meshes contain $\num{723132}$ curved elements or $\num{370243584}$ DOFs and $\num{880308}$ curved elements or $\num{450717696}$ DOFs for the $\alpha=\SI{0}{\degree}$ and $\alpha=\SI{3}{\degree}$ test case respectively. The span-wise extension of the computational domain is set to $\Delta_z=0.05c$ and periodic boundary conditions are used in $z$-direction. The airfoil itself uses an isothermal wall in order to mimic the very short experimental measurement times due to the vacuum-type indraft design of the wind tunnel (\cf \sref{sec:ExpApp}).

Due to a lack of data availability, we neglected the influence of incoming inflow turbulence in the wind tunnel and thus used a Dirichlet type free-stream boundary condition at the inflow. The outflow boundary is realized using a subsonic outflow condition based on the Mach number \cite{Carlson:2011}. The flow is sponged after inflow and before outflow to damp the reflections of the acoustics in the wind tunnel. \section{Results\label{sec:results}}

\begin{figure}[tb!]
\centering
\begin{subfigure}[t]{0.48\textwidth}
\centering
\includegraphics[width=0.9\columnwidth]{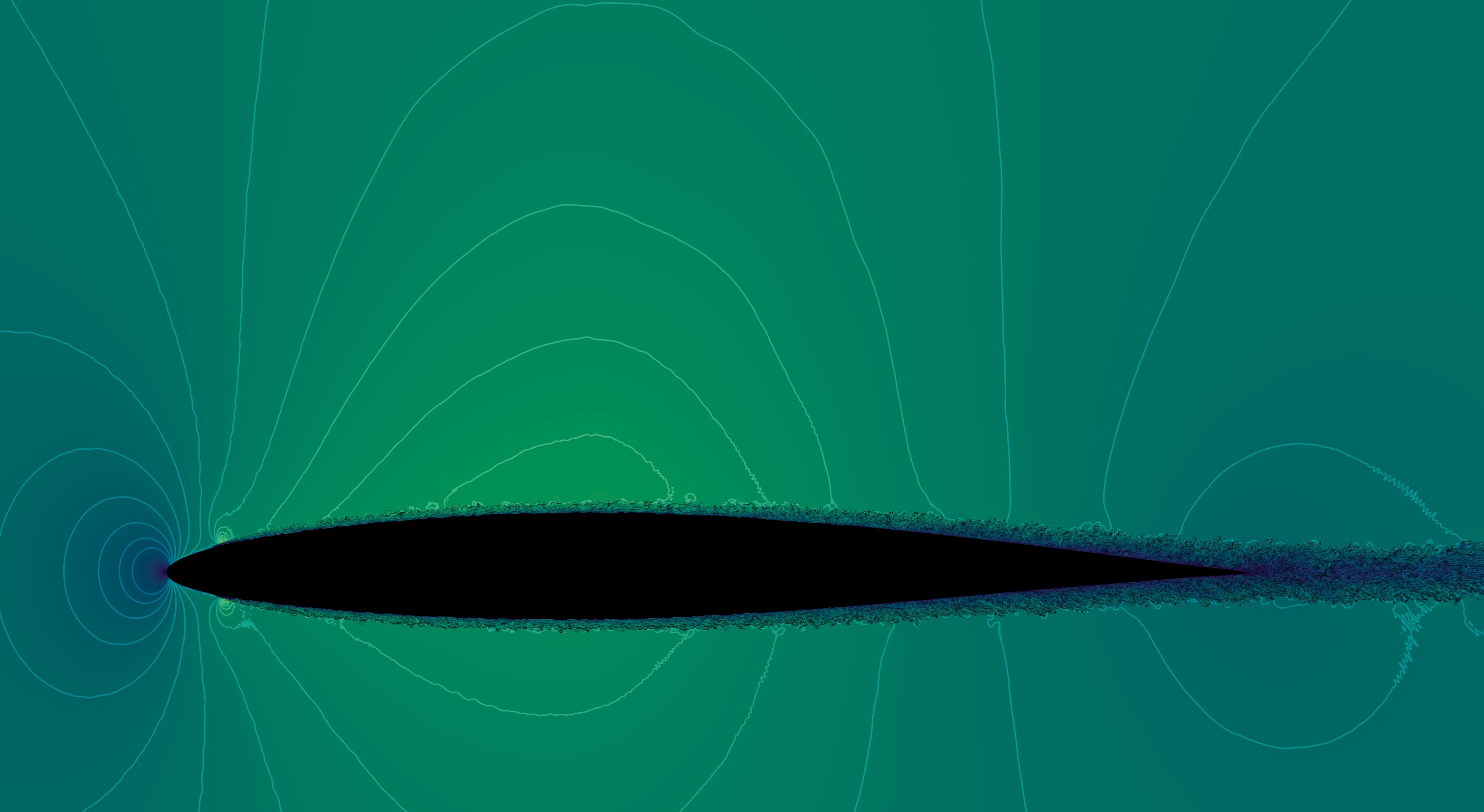}
\caption{Instantaneous flow field for the $\alpha=\SI{0}{\degree}$ simulation.}
\label{fig:MaQCrit_AoA0p0}
\end{subfigure}
\hfill
\begin{subfigure}[t]{0.48\textwidth}
\centering
\includegraphics[width=0.9\columnwidth]{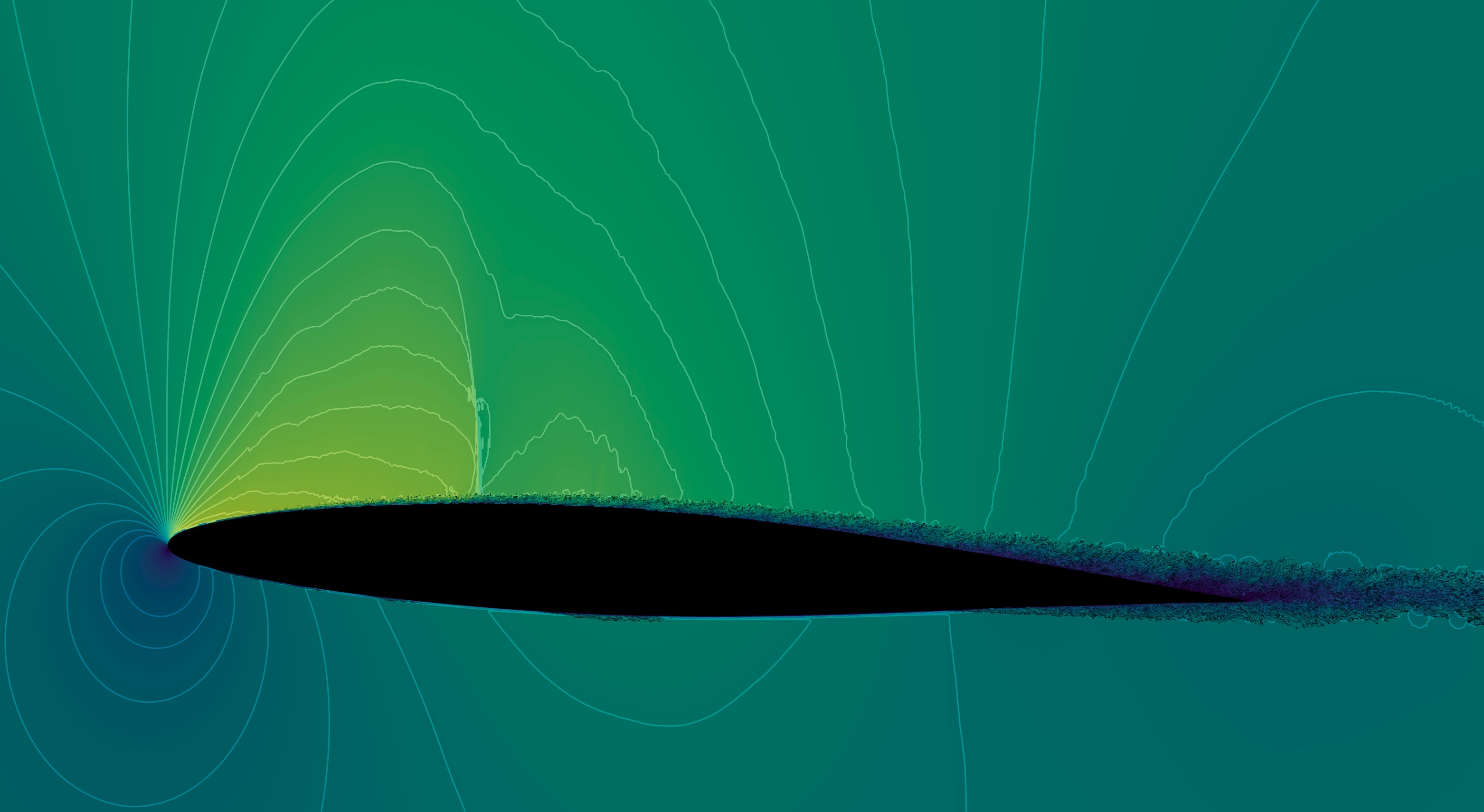}
\caption{Instantaneous flow field for the $\alpha=\SI{3}{\degree}$ simulation.}
\label{fig:MaQCrit_AoA3p0}
\end{subfigure}
\includegraphics[width=0.5\columnwidth]{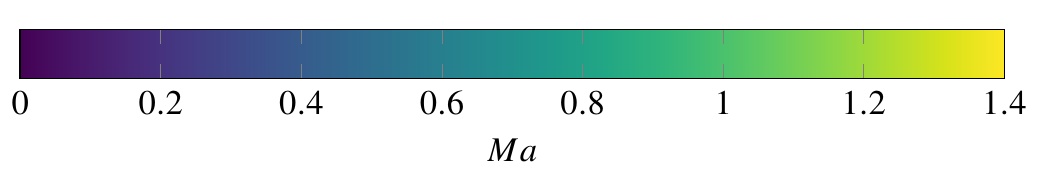}
\caption{Comparison of the instantaneous solution of the wall-resolved large eddy simulation containing iso-contour lines at fixed $Ma$-numbers. Iso-surfaces of the Q-Criterion are used to visualize turbulent structures.}
\label{fig:MaQCrit}
\end{figure}

Now we evaluate and compare the results of the PIV measurements and the WRLES. In \fref{fig:MaQCrit} the WRLES instantaneous flow field for both test cases is visualized showing differences in boundary layer thickness, Mach number and thus the presence of a shock in the $\alpha=\SI{3}{\degree}$ test case.

The color map and the iso-contour lines are matched between the two simulations. For $\alpha=\SI{0}{\degree}$ we can see that the Mach number reaches $Ma=1$ at a maximum, while in case of $\alpha=\SI{3}{\degree}$ we have a significant supersonic area on the suction side of the airfoil. Thus, we expect the $\alpha=\SI{3}{\degree}$ test case to be the more challenging one for comparing experimental and numerical data. Additionally, comparing the iso-contour lines we can see, that the uninclined airfoil shows a nearly symmetric behavior due to its relatively low curvature.

The evaluation for both test cases ($\alpha=\SI{0}{\degree}$ \& $\alpha=\SI{3}{\degree}$) will follow the same scheme. We first compare the mean velocity components $\overline{u}$ and $\overline{v}$, followed by assessing the difference in the mean velocity fluctuations $\overline{u'u'}$, $\overline{v'v'}$ and $\overline{u'v'}$. We always show the relative or absolute error of the simulation data with respect to the experimental PIV data. We conclude by showing more detailed plots of the flow field on the suction side and a $c_f^*$ skin friction type plot generated from the PIV data.

In order to compare the PIV and the WRLES results, the data had to be calibrated due to small deviations in the reference system. Hence, the coordinate system of the experimental data was translated and rotated. This was necessary since the PIV data is calibrated with respect to the wall. However, due to limitations arising from laser reflections on the model surface the flow field cannot be completely captured in the region very close to the wall. To compensate this, the coordinate system of the experimental data was shifted by $2\%c$ in $x$- and $1.5\%c$ in $y$-direction. To find these values we used an optimization algorithm to match the position of the leading edge as well as the center line of the wake. For the $\alpha=\SI{3}{\degree}$ test case the coordinate system additionally was rotated by $\SI{0.3}{\degree}$ in order to compensate small deviations in the reference system and thus to match the free stream velocity components.

\subsection{Test Case 1: $\alpha=\SI{0}{\degree}$\label{sec:AoA0p0}}

\begin{figure}[t]
\centering
\begin{subfigure}[t]{0.48\textwidth}
\centering
\includegraphics{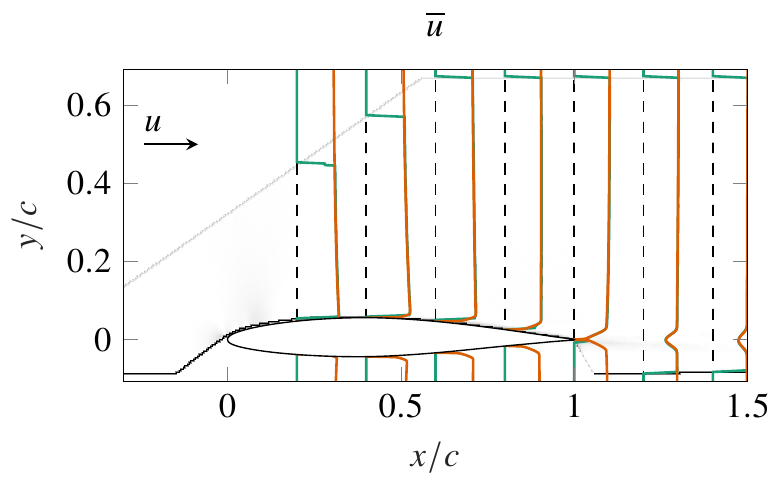}	
\end{subfigure}
\hfill
\begin{subfigure}[t]{0.48\textwidth}
\centering
\includegraphics{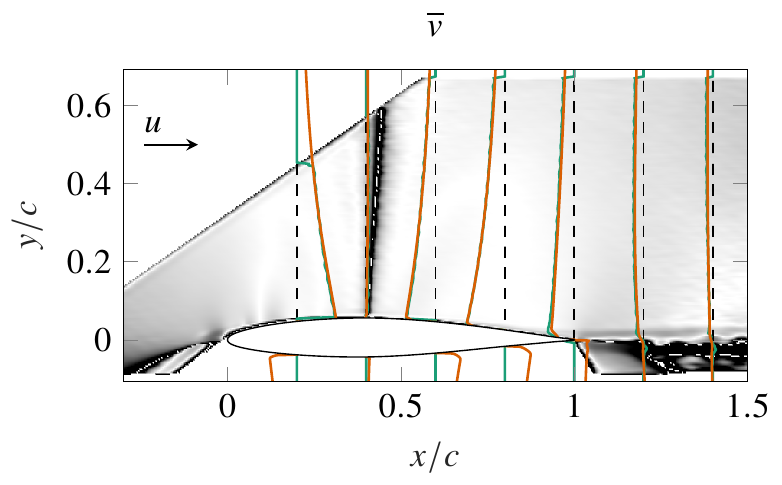}
\end{subfigure}
\vskip\baselineskip
\begin{subfigure}[t]{0.48\textwidth}
\centering
\includegraphics{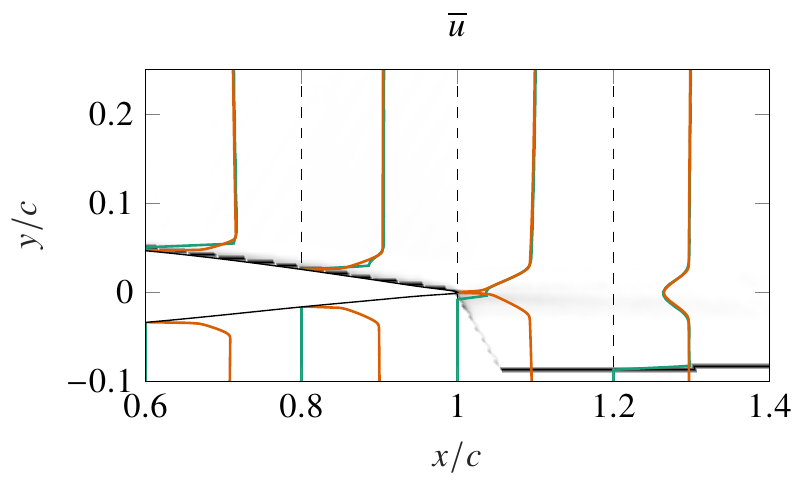}
\end{subfigure}
\hfill
\begin{subfigure}[t]{0.48\textwidth}
\centering
\includegraphics{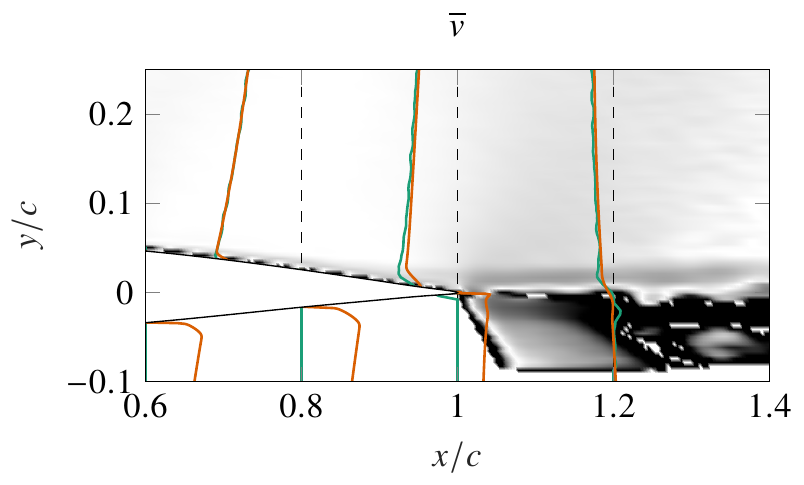}
\end{subfigure}
\begin{minipage}{0.4\columnwidth}
\centering
\includegraphics{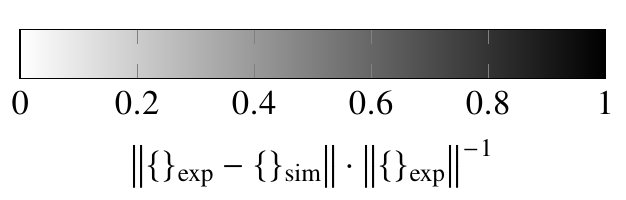}
\end{minipage}
\hspace{0.05\columnwidth}
\begin{minipage}{0.4\columnwidth}
\centering
\includegraphics{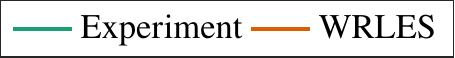}
\end{minipage}
\centering
\caption{Comparison of the mean velocity components for $\alpha=\SI{0}{\degree}$ between PIV and WRLES.}
\label{fig:AoA0p0_mean_comp}
\end{figure}

We start by evaluating the reference $\alpha=\SI{0}{\degree}$ test case. In \fref{fig:AoA0p0_mean_comp} we can see the comparison of the mean velocity components $\overline{u}$ and $\overline{v}$ for the entire field of view of the camera (\cf \fref{fig:AoA3p0_comp_inst}). In the background the relative error of the simulation data compared to the experimental results is plotted. The color map and the range are equivalent for the $\overline{u}$ and $\overline{v}$ plot. Additionally, each of the plots contains not only the information about the relative error, but also extracted velocity profiles at the black dashed lines. Those are shown for WRLES and PIV and allow for even better comparison of the data. The extraction lines are placed each $\Delta x/c=0.2$ starting at $x/c=0.2$.

Especially the $\overline{u}$-plots show two important results. First, the boundary layer profiles match very well also close to the trailing edge, where we have a thicker boundary layer and thus the PIV system is capable of capturing a greater part of the boundary layer. Second, the wake area is also captured very well and shows that the resulting wake has the same dimensions and velocity magnitudes in the PIV and the WRLES. The vertical velocity component $\overline{v}$ also shows good agreement, but in parts a greater relative error compared to the $\overline{u}$-component. A detailed analysis of the RMS-error along the extracted boundary layer profile can be found in \tref{tab:AoA0p0_RMS}. The error is calculated using the mean velocity data extracted at the vertical dashed black lines in \fref{fig:AoA0p0_mean_comp}. We calculate the RMS-error for all data points on those lines where data is available for both the WRLES and the PIV measurements. This is crucial since the PIV data is not available all the way to the wall. The greater relative error of the $\overline{v}$ component is since the sign of this velocity component changes and thus the relative error can become greater close to $\overline{v}\approx0$. The velocity profiles on the airfoil show good agreement between WRLES and PIV. However, close to the trailing edge and in the wake, we can see that the experiment shows slightly greater acceleration in $y$ and thus a bigger difference ($17.74\%$ at $x=1c$) between the positive and negative $v$-velocities in the wake.

\begin{figure}[t]
\centering
\begin{subfigure}[t]{0.32\textwidth}
\centering
\includegraphics{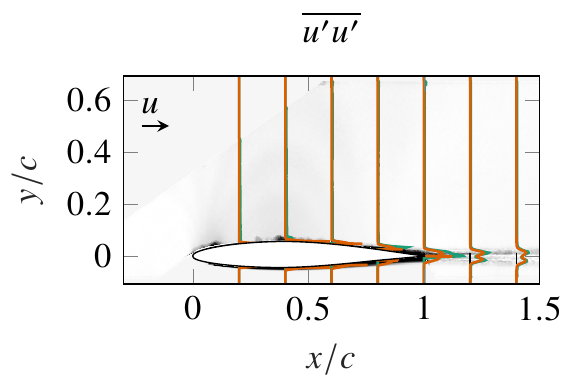}
\end{subfigure}
\hfill
\begin{subfigure}[t]{0.32\textwidth}
\centering
\includegraphics{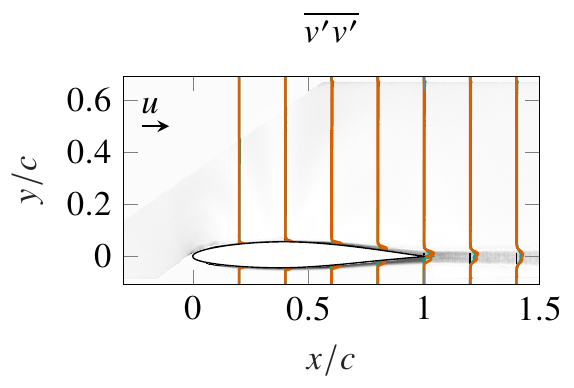}
\end{subfigure}
\hfill
\begin{subfigure}[t]{0.32\textwidth}
\centering
\includegraphics{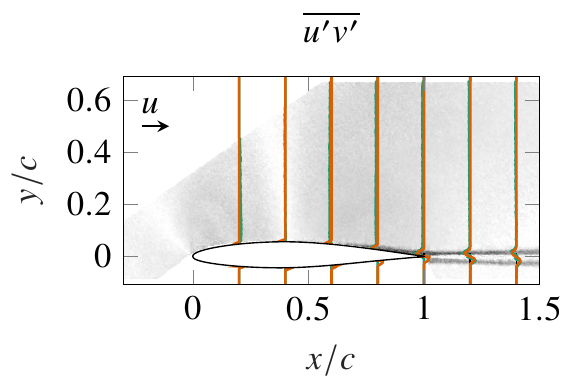}
\end{subfigure}
\vskip\baselineskip
\begin{subfigure}[t]{0.32\textwidth}
\centering
\includegraphics{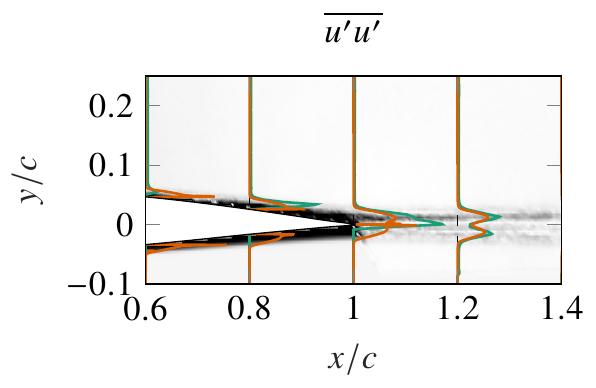}
\end{subfigure}
\hfill
\begin{subfigure}[t]{0.32\textwidth}
\centering
\includegraphics{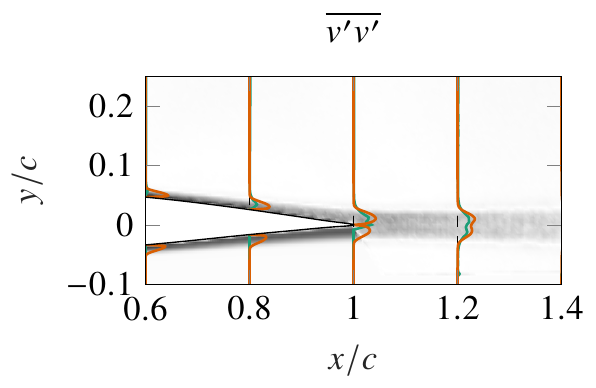}
\end{subfigure}
\hfill
\begin{subfigure}[t]{0.32\textwidth}
\centering
\includegraphics{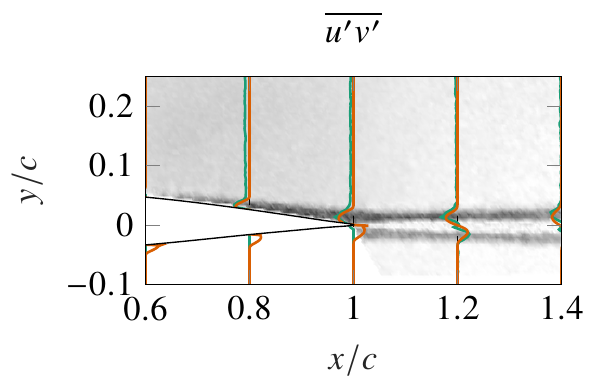}
\end{subfigure}
\begin{minipage}{0.4\columnwidth}
\centering
\includegraphics{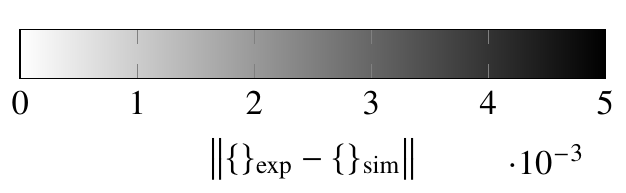}
\end{minipage}
\hspace{0.05\columnwidth}
\begin{minipage}{0.4\columnwidth}
\centering
\includegraphics{figures/legend}
\end{minipage}
\centering
\caption{Comparison of the mean velocity fluctuation components for $\alpha=\SI{0}{\degree}$ between PIV and WRLES.}
\label{fig:AoA0p0_fluc_comp}
\end{figure}

\begin{table}[t]
\caption{RMS-errors of the boundary layer sections of the $\alpha=\SI{0}{\degree}$ test case. The error is calculated on the suction side of the airfoil and in the wake area.}
\label{tab:AoA0p0_RMS}
\centering
\begin{tabular}{lllllllll}
\hline
RMS-Error & $x=0.2c$     & $x=0.4c$     & $x=0.6c$     & $x=0.8c$     & $x=1c$       & $x=1.2c$     & $x=1.4c$     & Mean    \\ \hline
$\overline{u}$          & 0.95\%	& 0.67\%	& 1.03\%	& 1.19\%	& 1.15\%	& 1.00\%	& 0.87\%	& 0.98\%   \\
$\overline{v}$          & 3.50\%	& 46.31\%	& 3.60\%	& 0.82\%	& 7.89\%	& 17.74\%	& 24.47\%	& 14.90\%  \\
\hline
\end{tabular}\end{table}

We continue by investigating the velocity fluctuations which are visualized in \fref{fig:AoA0p0_fluc_comp}. From the instantaneous PIV components $u$ and $v$, we calculate the fluctuations $\overline{u'u'}$, $\overline{v'v'}$ and $\overline{u'v'}$. The calculation of the fluctuations for the experimental data uses the same algorithms that are also used for the WRLES data such that we can ensure comparability of the results. This time we do not plot the relative, but the absolute error. This is due to the fact, that the experimental fluctuations do not completely vanish in the free stream and stay in the range of $\mathcal{O}(10^{-4})$ due to the low but inevitable turbulent level of the wind-tunnel flow. The numerical data, however, reach values that are orders of magnitude smaller. Another reason is that for calculating the fluctuations we only have $\mathcal{O}(10^3)$ samples while in WRLES we utilize $\mathcal{O}(10^5)$ samples. Thus, we adjust the mean value of the WRLES fluctuations to the PIV mean values to ensure comparability. The colormaps again are matched for all plots in \fref{fig:AoA0p0_fluc_comp}.

Comparing the results in \fref{fig:AoA0p0_fluc_comp} we can see that matching the fluctuations is more difficult. The $\overline{u'u'}$ component this time shows good agreement in the wake and on the surface of the airfoil. Again, the limitations of the PIV system limit the comparability very close to the airfoil due to reflections of the laser light sheet on the surface as well as seeding of the particles in the wake area. Especially in high-speed flows the PIV particles can be transported out of the PIV plane. We can see that the WRLES underestimates the fluctuations in comparison with the experiment. This is especially visible in the wake area and close to the trailing edge, where the fluctuations resulting from the pressure side seem to be sufficiently captured, but the contribution from the suction side in the wake area is slightly underestimated resulting in a more symmetric $\overline{u'u'}$ distribution in the WRLES.

Also the components $\overline{v'v'}$ and $\overline{u'v'}$ show good agreement on the airfoil and in the wake area. The WRLES and the PIV are able to capture the slight asymmetry resulting from the non-symmetric airfoil. Also close to the trailing edge where the boundary layer thickens up, we can see that WRLES $\overline{u'v'}$ values match the experiment very well. Additionally, in the wake the inflection point of the fluctuations is matched. For $\overline{v'v'}$ we qualitatively see the same results. The only difference there is a slight overestimation of the WRLES in the fluctuations in $y$-direction compared to the experimental data.

This general small mismatch between experiment and simulation can result from the limitations of the experiment mentioned above. Due to the sampling frequency of the camera, we do not have as many snapshots for calculating the fluctuations. Additionally, background turbulence resulting from the inflow in the channel can further affect the velocity fluctuations and prevent them from reaching a value as small as the simulation.

\begin{figure}[t]
\centering
\includegraphics{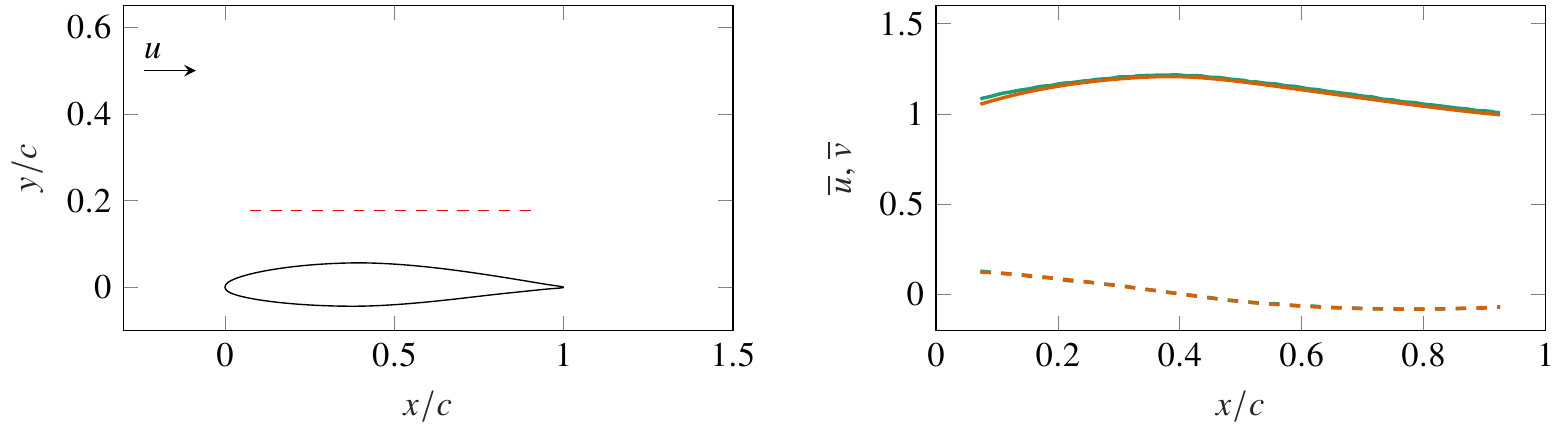}
\includegraphics{figures/legend}
\hspace{0.05\columnwidth}
\includegraphics{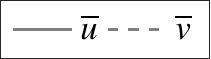}
\caption{Mean velocity components evaluated at a section above suction side of airfoil.}
\label{fig:AoA0p0_section_vel_mean}
\end{figure}

We continue by investigating the mean velocity components in stream wise direction. The results are visualized in \fref{fig:AoA0p0_section_vel_mean}. On the left part of the figure, we see the airfoil and a red dashed extraction line. To generate the plots on the right side we extracted the values of the flow field along this line. The results are plotted in the right part of \fref{fig:AoA0p0_section_vel_mean}. Again, we can see good agreement for both velocity components underlining the observations made in \fref{fig:AoA0p0_mean_comp}. $\overline{u}$ as well as $\overline{v}$ also show little difference between WRLES and experiment in the weak low pressure area on the suction side (\cf \fref{fig:MaQCrit}) and close to the trailing edge of the airfoil.

\begin{figure}[t]
\centering
\includegraphics{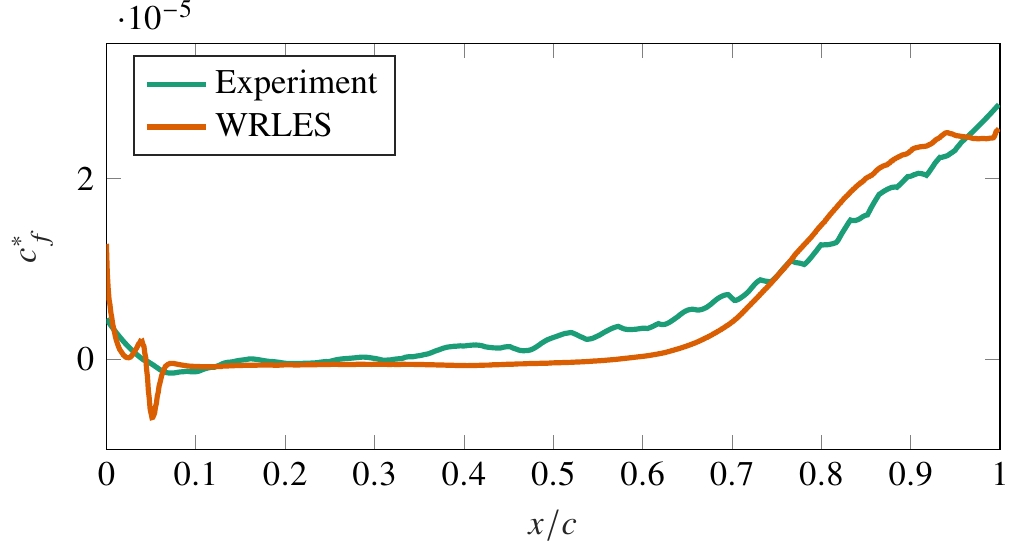}
\caption{Pseudo skin friction coefficient evaluated near the wall to compare gradients.}
\label{fig:AoA0p0_cf}
\end{figure}

Finally, we also want to compare the velocity gradient in the solution. Optimally, one would want to compare the skin friction at the wall for both approaches. However, since the PIV system is not capable of capturing the data sufficiently close to the wall due to the reasons discussed above, we cannot use the typical approach by taking the gradient directly at the wall. Typically, the lower 10\%-15\% of the boundary layer cannot be captured by the PIV system. Instead, we evaluate the gradients on a virtual surface placed in the flow with an offset to the suction side of the airfoil.  We choose the offset such that both PIV and WRLES have data available at this distance. On this surface, we then evaluate a pseudo skin friction coefficient $c_f^\ast$. The result is visualized in \fref{fig:AoA0p0_cf}.

To generate \fref{fig:AoA0p0_cf} we extracted the mean velocities with an offset normal to the wall of $\Delta n \approx 0.0175c$. Overall, the results show good agreement and also match very well close to the leading and trailing edge of the airfoil. In the WRLES we can see a spike in $c_f^\ast$ close to $x/c=0.05$. This is due to the geometric tripping used in the $\alpha=\SI{0}{\degree}$ test case. This plot again underlines the good agreement in the mean values and proves that also the gradients close to the wall match very well.

\subsection{Test Case 2: $\alpha=\SI{3}{\degree}$\label{sec:AoA3p0}}

Next, we compare the results of the $\alpha=\SI{3}{\degree}$ test case. This angle of attack poses more challenges to the overall comparison between simulation and experiment, since the presence of a shock and the higher pressure gradients have a great effect on the flow.

\begin{figure}[t]
\centering
\begin{subfigure}[t]{0.48\textwidth}
\centering
\includegraphics{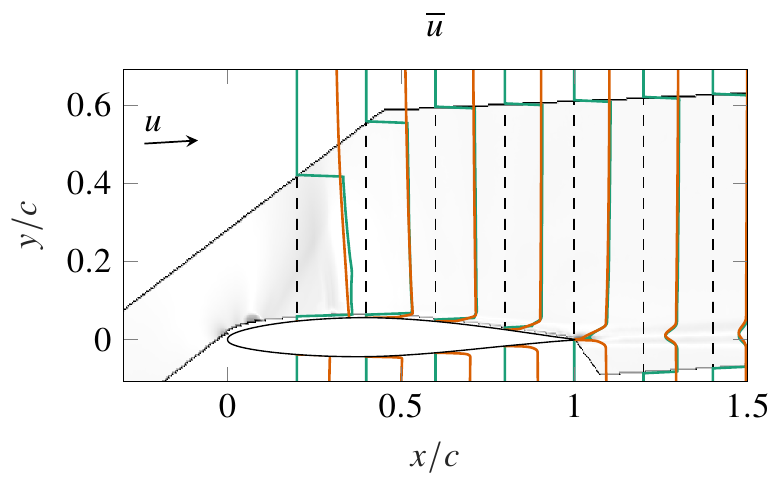}	
\end{subfigure}
\hfill
\begin{subfigure}[t]{0.48\textwidth}
\centering
\includegraphics{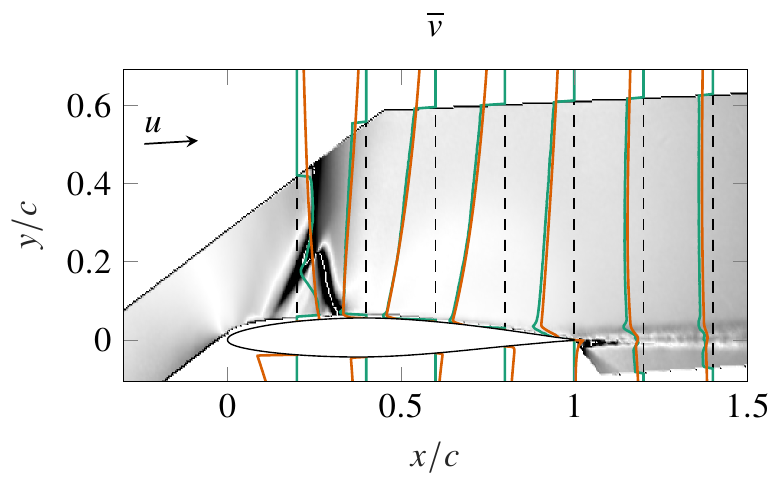}	
\end{subfigure}
\vskip\baselineskip
\begin{subfigure}[t]{0.48\textwidth}
\centering
\includegraphics{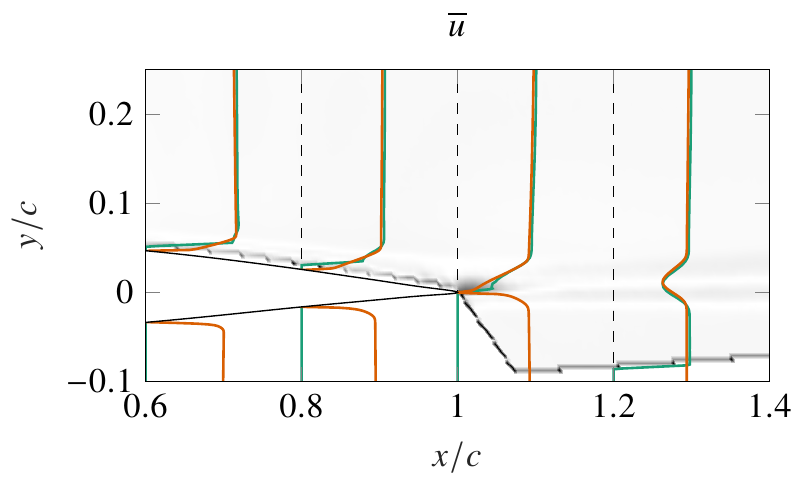}	
\end{subfigure}
\hfill
\begin{subfigure}[t]{0.48\textwidth}
\centering
\includegraphics{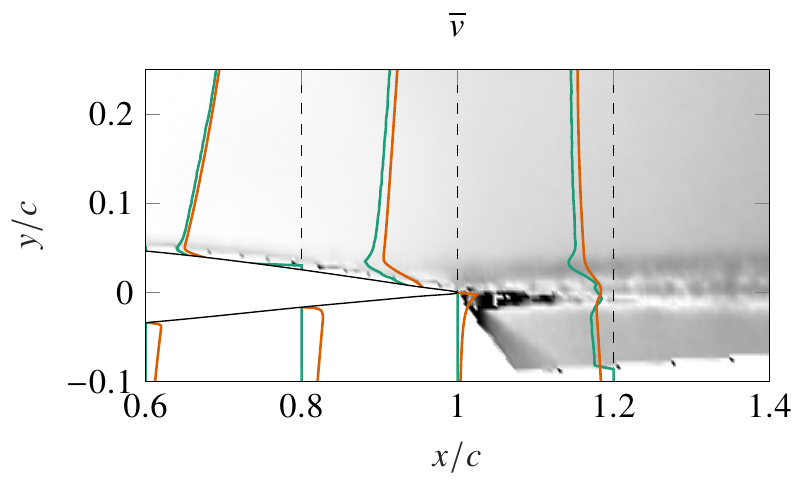}	
\end{subfigure}
\begin{minipage}{0.4\columnwidth}
\centering
\includegraphics{figures/mean_error_legend}
\end{minipage}
\hspace{0.05\columnwidth}
\begin{minipage}{0.4\columnwidth}
\centering
\includegraphics{figures/legend}
\end{minipage}
\centering
\caption{Comparison of the mean velocity components for $\alpha=\SI{3}{\degree}$ between PIV and WRLES.}
\label{fig:AoA3p0_mean_comp}
\end{figure}

We again start by evaluating the comparison of the mean velocity components $\overline{u}$ and $\overline{v}$ and plot the relative error in \fref{fig:AoA3p0_mean_comp}. Like for the  $\alpha=\SI{0}{\degree}$ test case, the $\overline{u}$-component of the WRLES and the experiment match very well on the airfoil and in the wake region. However, there is a light gray area at $x/c\in[0.1;0.3]$ on the suction side of the airfoil. In this region the steady shock is located. If we compare the $\overline{v}$-component of the flow field, we can see clearly that this area is also troubled and the experiment does not match the WRLES exactly. First, we can see that the main normal shock is slightly shifted. We are going to quantify the shift of the shock later. Second, there is another oblique structure visible which is not present in WRLES. If we additionally compare the instantaneous flow fields in \fref{fig:AoA3p0_comp_inst} we can observe that the difference in the shock structure is also present there. In the experiment we can see a lambda shaped shock, while the WRLES only shows one normal shock. Due to the difference in Mach number in front of the shock wave, we can also state that the normal shock in the WRLES is weaker. In preparation for this simulation, we also conducted inviscid Euler simulations which have shown that the shock in free stream conditions without the presence of wind tunnel walls is weaker and smaller. Thus, we argue that the higher intensity and the presence of the oblique shock wave in the experiment is caused by an additional displacement induced by the wind tunnel boundary layer. Especially, for low Reynolds number flows displacement effects have a severe influence on the flow physics (\cf \cite{Emmons:1948}). An additional displacement such as posed by the turbulent boundary layer at the wall is thus a plausible reason for the stronger shock structures in the experiment.

\begin{figure}[t]
\centering
\begin{subfigure}[t]{0.32\textwidth}
\centering
\includegraphics{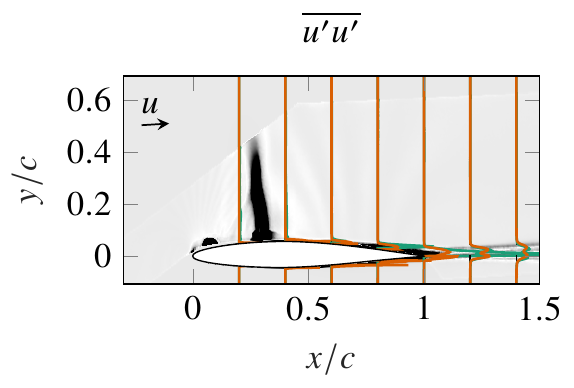}
\end{subfigure}
\hfill
\begin{subfigure}[t]{0.32\textwidth}
\centering
\includegraphics{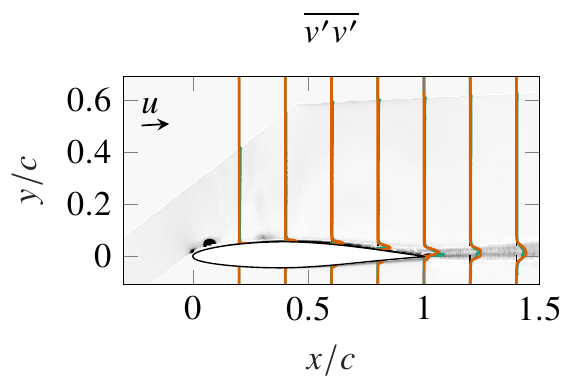}
\end{subfigure}
\hfill
\begin{subfigure}[t]{0.32\textwidth}
\centering
\includegraphics{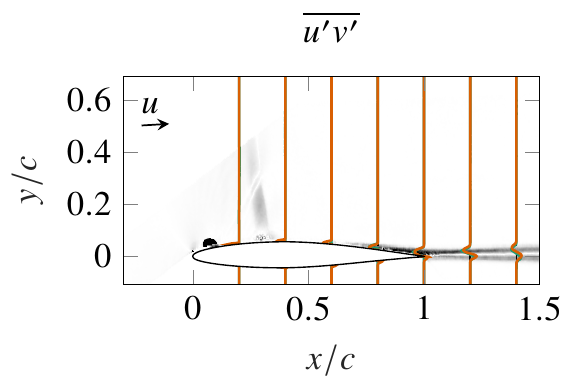}
\end{subfigure}
\vskip\baselineskip
\begin{subfigure}[t]{0.32\textwidth}
\centering
\includegraphics{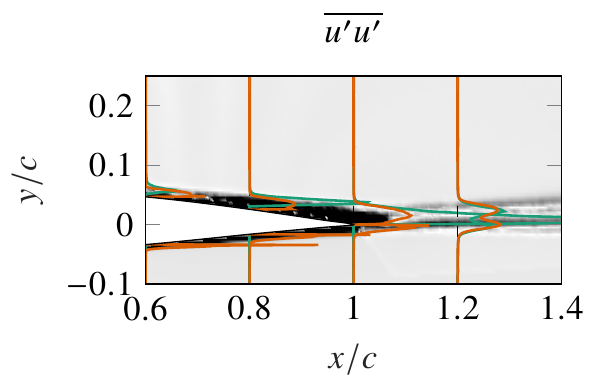}
\end{subfigure}
\hfill
\begin{subfigure}[t]{0.32\textwidth}
\centering
\includegraphics{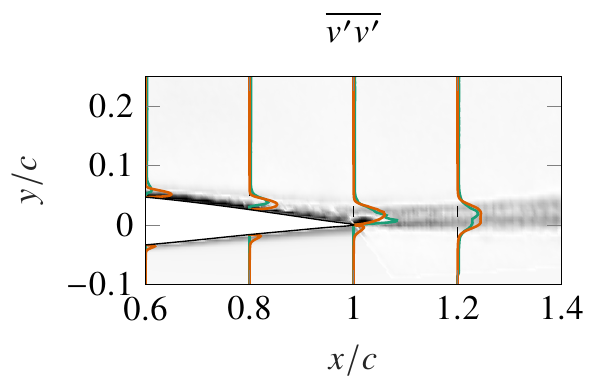}
\end{subfigure}
\hfill
\begin{subfigure}[t]{0.32\textwidth}
\centering
\includegraphics{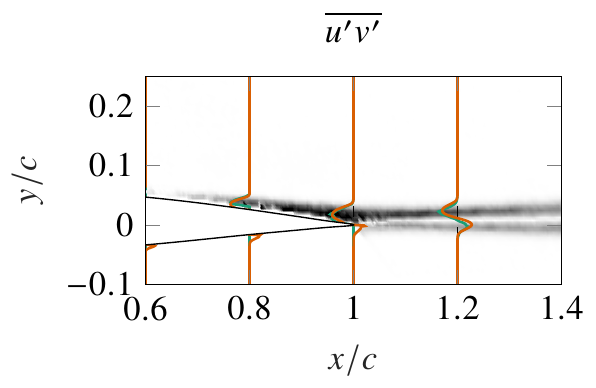}
\end{subfigure}
\begin{minipage}{0.4\columnwidth}
\centering
\includegraphics{figures/fluc_error_legend}
\end{minipage}
\hspace{0.05\columnwidth}
\begin{minipage}{0.4\columnwidth}
\centering
\includegraphics{figures/legend}
\end{minipage}
\centering
\caption{Comparison of the mean velocity fluctuation components for $\alpha=\SI{3}{\degree}$ between PIV and WRLES.}
\label{fig:AoA3p0_fluc_comp}
\end{figure}

\begin{table}[t]
\caption{RMS-errors of the boundary layer sections of the $\alpha=\SI{3}{\degree}$ test case. The error is calculated on the suction side of the airfoil and in the wake area.}
\label{tab:AoA3p0_RMS}
\centering
\begin{tabular}{lllllllll}
\hline
RMS-Error & $x=0.2c$     & $x=0.4c$     & $x=0.6c$     & $x=0.8c$     & $x=1c$       & $x=1.2c$     & $x=1.4c$     & Mean    \\ \hline
$\overline{u}$          & 8.54\%	& 0.55\%	& 2.55\%	& 3.09\%	& 3.23\%	& 3.40\%	& 3.57\%	& 3.56\%   \\
$\overline{v}$          & 69.10\%	& 9.68\%	& 4.35\%	& 6.39\%	& 14.97\%	& 24.98\%	& 33.95\%	& 23.35\%  \\
\hline
\end{tabular}\end{table}

We continue by comparing the mean velocity fluctuations $\overline{u'u'}$, $\overline{v'v'}$ and $\overline{u'v'}$ which are visualized in \fref{fig:AoA3p0_fluc_comp}. Again, we see good agreement between the experiment and the WRLES including low absolute errors. The results follow the observations made for the $\alpha=\SI{0}{\degree}$ test case seen in \fref{fig:AoA3p0_fluc_comp}. For $\overline{v'v'}$ the fluctuations resulting from the WRLES are slightly overpredicting those from the experiment. Vice versa for $\overline{u'u'}$. Also  for $\overline{u'v'}$ the error is low and we see good agreement on the surface of the airfoil as well as in the wake area. A detailed overview over the RMS-errors of the boundary layer sections for the $\alpha=\SI{3}{\degree}$ test case can be found in \tref{tab:AoA3p0_RMS}.

In all the mean velocity fluctuations plots we can see a black spot close to the leading edge on the suction side of the airfoil. This is caused by the tripping tape used in the experiment, which caused reflections of the laser light sheet impinging on the airfoil surface. Thus, this does not affect the velocity field and is only a measurement artifact.

Earlier we described that we encountered difficulties tripping the incoming laminar boundary layer. Thus, the flow on the pressure side in the $\alpha=\SI{3}{\degree}$ test case transitions naturally in the numerical simulation. Since the mean velocity fluctuations generally match well it remains unclear whether the tripping in the experiment worked or if the flow there also relaminarized before naturally transitioning to turbulent flow close to the trailing edge of the airfoil.

\begin{figure}[t]
\centering
\begin{subfigure}[t]{\textwidth}
\begin{minipage}{0.48\textwidth}
\centering
\includegraphics[width=\columnwidth]{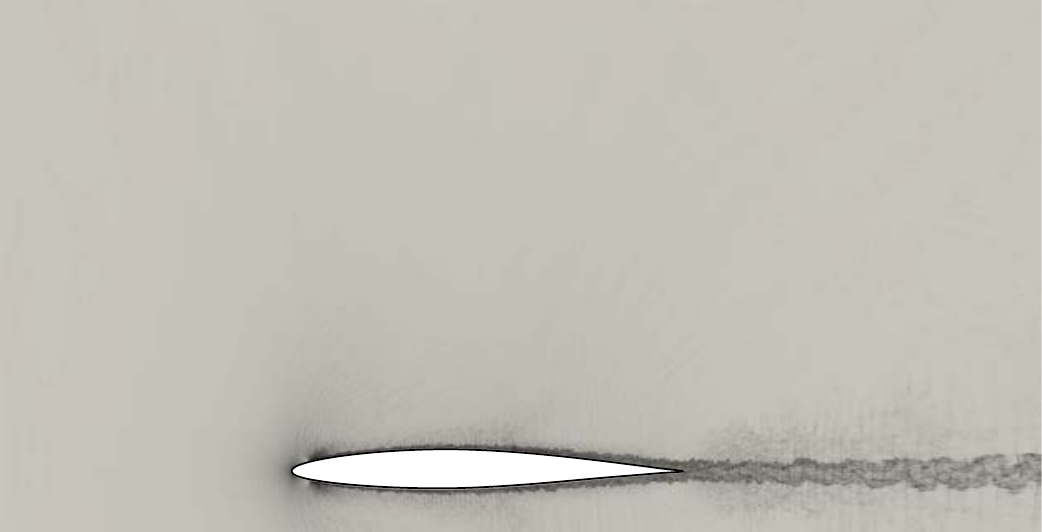}
\end{minipage}
\hfill
\begin{minipage}{0.48\textwidth}
\centering
\includegraphics[width=\columnwidth]{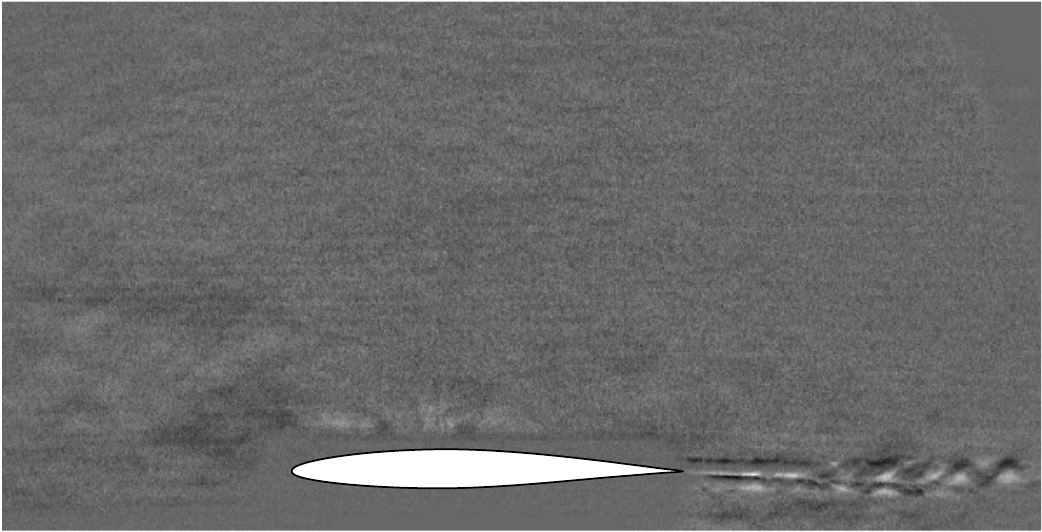}
\end{minipage}
\caption{Numerical results (left) and experimental results (right) for $\alpha=\SI{0}{\degree}$.}
\label{fig:schlieren_comp_AoA0p0}
\end{subfigure}
\vskip\baselineskip
\begin{subfigure}[t]{\textwidth}
\begin{minipage}{0.48\textwidth}
\centering
\includegraphics[width=\columnwidth]{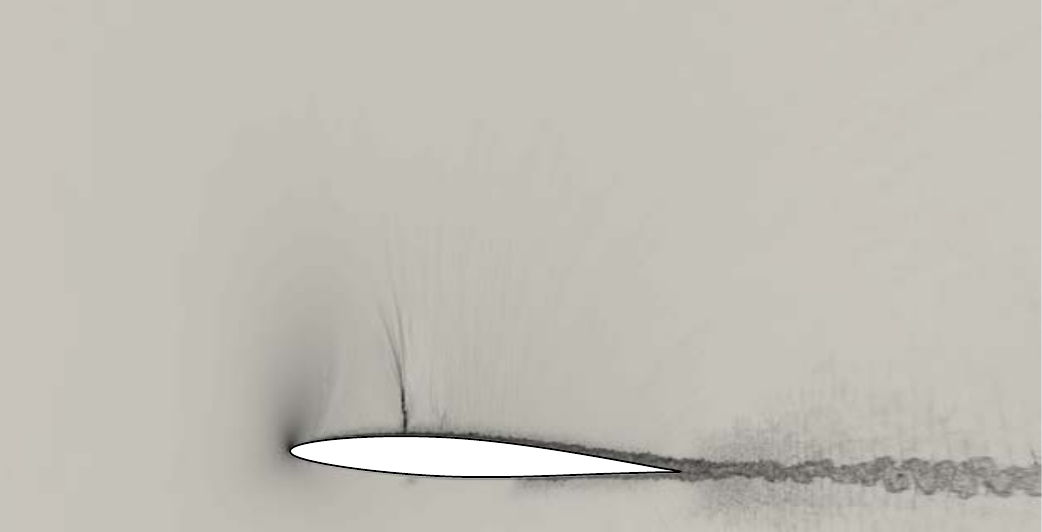}
\end{minipage}
\hfill
\begin{minipage}{0.48\textwidth}
\centering
\includegraphics[width=\columnwidth]{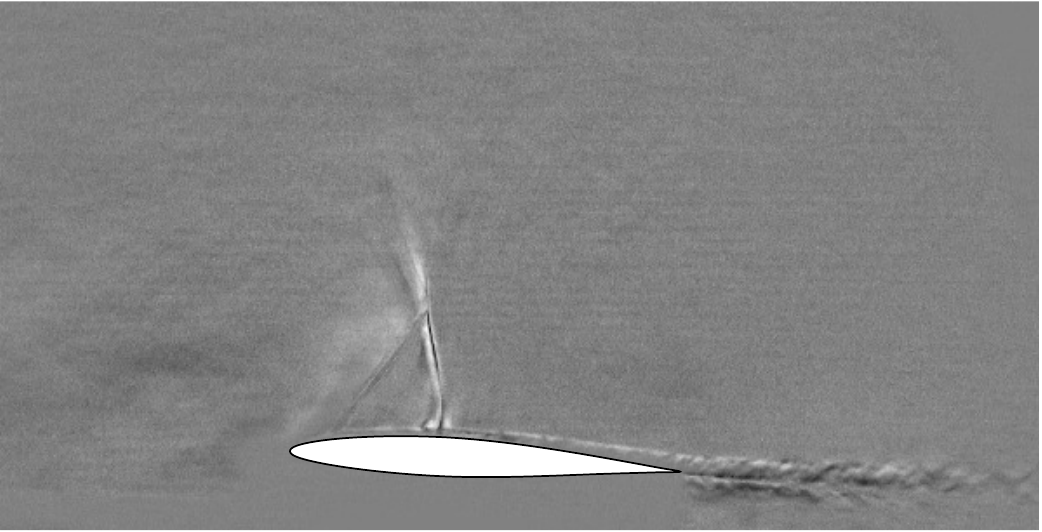}
\end{minipage}
\caption{Numerical results (left) and experimental results (right) for $\alpha=\SI{3}{\degree}$.}
\label{fig:schlieren_comp_AoA3p0}
\end{subfigure}
\centering
\caption{Exemplary focusing schlieren snapshots and numerical schlieren visualization of the shock-free $\alpha=\SI{0}{\degree}$ case (top), and steady shock case obtained for $\alpha=\SI{3}{\degree}$ (bottom).}
\label{fig:schlieren_comp}
\end{figure}

The assumption of a naturally transitioning pressure side is supported by observations in the region downstream of the trailing edge, where suction and pressure side structures coalesce. Schlieren representations of both the $\alpha=\SI{0}{\degree}$ and $\alpha=\SI{3}{\degree}$ cases (see \fref{fig:schlieren_comp}) show a clear dividing line between the strongly perturbed upper half of the wake and the substantially more even flow in the pressure side wake. Alternating bright and dark spots indicate vortical structures. These spots are visible all along the upper part of the captured wake (originating from the suction side), whereas turbulent structures in the lower part (stemming from the pressure side) dissipate much earlier, i.e. within $x=0.5c$ from the trailing edge. The latter observation suggests a less turbulent nature of the pressure-side boundary layer. For comparison the numerical schlieren representation of two instantaneous snapshots  is also added to \fref{fig:schlieren_comp}. While not being fully comparable they show good agreement of the vortical structures in the wake and the overall boundary layer thickness close to the trailing edge of the airfoil at both angles of attack.

\begin{figure}[t]
\centering
\includegraphics{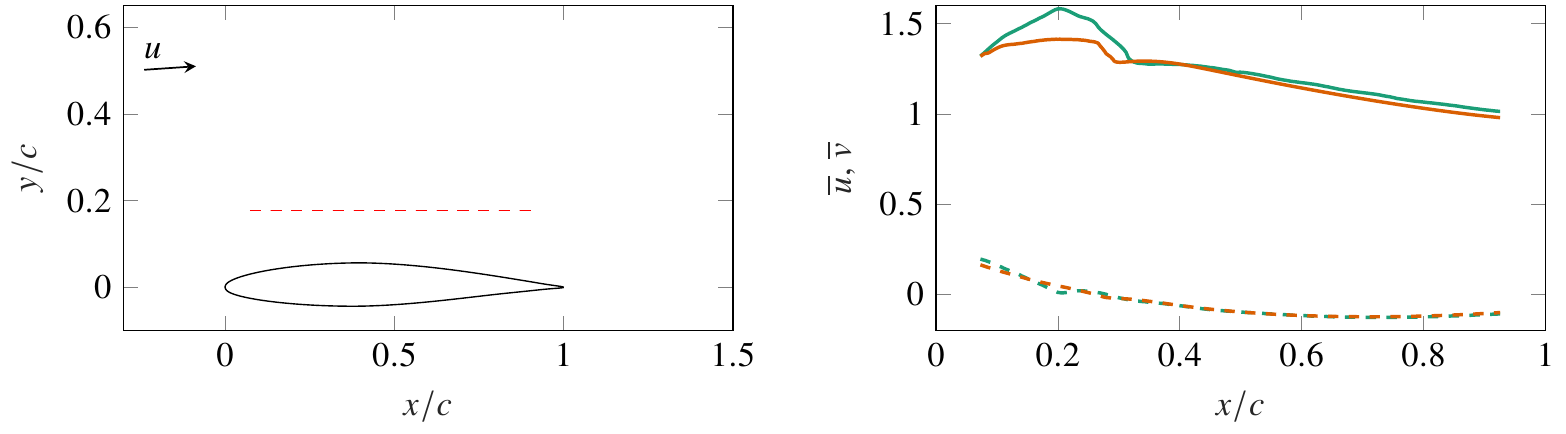}
\includegraphics{figures/legend}
\hspace{0.05\columnwidth}
\includegraphics{figures/section_legend}
\caption{Mean velocities evaluated at section above suction side of airfoil.}
\label{fig:AoA3p0_section_vel_mean}
\end{figure}

We continue evaluating the $\alpha=\SI{3}{\degree}$ test case by comparing the mean velocity components in stream parallel direction. The results are plotted in \fref{fig:AoA3p0_section_vel_mean} and again show the extraction line in the left part of \fref{fig:AoA3p0_section_vel_mean} and the extracted mean velocity components on the right side of the same figure.

Especially the $\overline{u}$-component shows important aspects of the flow in the low pressure area on the suction side. First, we now can quantify the exact position of the main normal shock wave, which is located at $x/c=0.30$ in the experiment and at $x/c=0.28$ in the WRLES, yielding a difference of only $\si{2\percent}c$. Second, we can see the higher $\overline{u}$ levels in front of the shock in the experimental data underlining the observations made in \fref{fig:AoA3p0_comp_inst} which suggested that the shape differs and the intensity of the shock is higher in the experimental results. Third, the flow after the shock is also affected by the difference in strength of the shock, resulting in overall lower velocities directly behind the shock in the PIV data (\cf \fref{fig:AoA3p0_section_vel_mean}).

\begin{figure}[t]
\centering
\includegraphics{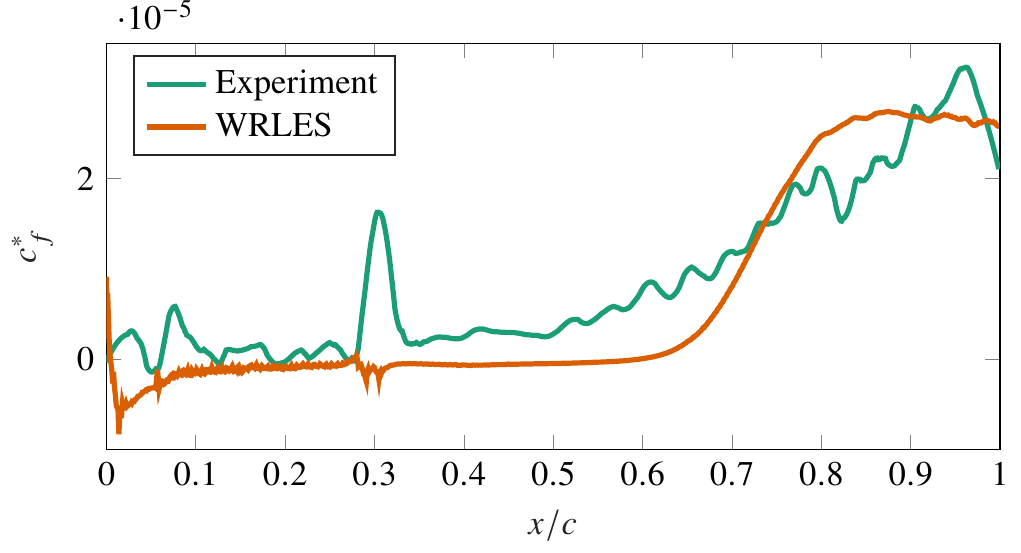}
\caption{Pseudo skin friction coefficient evaluated near the wall to compare gradients.}
\label{fig:AoA3p0_cf}
\end{figure}

Finally, we compare the results of the skin friction style $c_f^\ast$ shown in \fref{fig:AoA3p0_cf}. Again, the gradient was constructed at a virtual wall-normal plane at which both WRLES and experiment provide sufficient data. For the $\alpha=\SI{3}{\degree}$ test case this could be done for $\Delta n \approx 0.0175c$. Close to the leading edge we see good agreement between experiment and simulation. At the trailing edge the WRLES also matches the experimental results. In between we see a short deviation which can be explained with the slightly different flow field induced by the two shock shapes. However, there are two significant properties visible. First, we can validate the results of the shock position obtained from \fref{fig:AoA3p0_section_vel_mean}. The $c_f^\ast$ plot shows the shock at the same position and also provides information about the change in $\overline{u}$-velocity in wall-normal direction. Since the jump in $c_f^\ast$ is much larger in the experimental data we can conclude that the wall-normal change in velocity over the shock is much larger in the experiment.

Furthermore, we can observe that the numerical tripping method in the simulation does not produce as much noise as the geometric trip used in the $\alpha=\SI{0}{\degree}$ test case. This underlines the observations made in \fref{fig:trips} and makes the numerical trip a suitable candidate for future airfoil simulations. This trip additionally can be modified without remeshing the airfoil.
\section{Conclusion\label{sec:conclusion}}

In this paper we introduced a carefully designed reference data set, consisting of both experimental and numerical data of a NACA 64A-110 airfoil at $Ma=\num{0.72}$ and $Re_c=\num{930000}$ for $\alpha=\SI{0}{\degree}$ and $\alpha=\SI{3}{\degree}$. The data set will enable numerical validation, feature development or wall modeling of complex transonic flows including shocks. The data was generated running a wall-resolved large eddy simulation and conducting wind tunnel experiments. Care was taken to ensure that during the setup of the simulation and the preparation of the experiments the parameters and the overall setups match closely. This is why in the simulation we also reproduced the exact wind tunnel wall contours that were used during the test campaign.

The results show good agreement in the entire flow field between PIV and WRLES data. The wake dimensions and its orientation also match closely. In this work, we additionally present test cases at different angles of attack in order to also compare the flow field for more complex phenomena such as the presence of a steady shock on the suction side of the airfoil. The shock location in the experiment agrees with the simulation, however, the structure of the shock was different. Experience with this wind tunnel suggests that this difference results from the presence of the boundary layer at the wind tunnel walls, which was not included in the simulation. Still, the flow field after the shocks was comparable and also showed very few deviations between experiment and simulation. Additionally, we introduced a skin friction style coefficient that compared the wall-normal gradient of the $\overline{u}$-velocity component. Also, for this comparison we showed good agreement between experiment and simulation.

Overall, it can be stated that although the experimental results of the $\alpha=\SI{3}{\degree}$ test case were captured well, the $\alpha=\SI{0}{\degree}$ test case generally shows lower differences. This however is expected since the challenges to numerics and the overall setup are lower there. The complexity is further reduced by the fact that no change of the wind tunnel wall geometry was necessary which additionally removes sources of errors.

In future applications we plan to compare the reference data with lower fidelity simulation such as wall-modeled LES and discuss if these methods are also capable of capturing all the physical properties of the flow field for both angles of attack.

\section*{Acknowledgments and Funding Sources}
The authors gratefully acknowledge the Deutsche Forschungsgemeinschaft DFG (German Research Foundation) for funding this work in the framework of the research unit FOR 2895, and thank the Gauss Centre for Supercomputing e.V. (\url{www.gauss-centre.eu}) for funding this project (GCS-lesdg) by providing computing time on the GCS Supercomputer HAWK at High-Performance Computing Center Stuttgart (\url{www.hlrs.de}). We also gratefully acknowledge the contribution and support of Nick Capellmann, Dennis Matysik and the entire workshop team of the Institute of Aerodynamics during the manufacturing process of the wind tunnel model.

\end{document}